\documentclass[twocolumn]{aastex63}
\pdfoutput=1
\usepackage{graphicx}
\usepackage{enumerate}
\usepackage{amssymb}
\usepackage{amsmath} 
\usepackage{bm}
\usepackage{natbib}
\usepackage{color}
\usepackage{hyperref}
\usepackage{xspace}
\newcommand{\codename}{{\tt orvara}\xspace}
\newcommand{\gaia}{{\it Gaia}\xspace}
\newcommand{\hipparcos}{{\it Hipparcos}\xspace}

\newcommand{\Msun}{\mbox{$M_{\sun}$}}

\newcommand{\Mjup}{\mbox{$M_{\rm Jup}$}}

\begin{document}

\title{\codename: An Efficient Code to Fit Orbits using Radial Velocity, Absolute, and/or Relative Astrometry}

\author[0000-0003-2630-8073]{Timothy D.~Brandt}
\affiliation{Department of Physics, University of California, Santa Barbara, Santa Barbara, CA 93106, USA}

\author[0000-0001-9823-1445]{Trent J.~Dupuy}
\affiliation{Institute for Astronomy, University of Edinburgh, Royal Observatory, Blackford Hill, Edinburgh, EH9 3HJ, UK}

\author[0000-0002-6845-9702]{Yiting Li}
\affiliation{Department of Physics, University of California, Santa Barbara, Santa Barbara, CA 93106, USA}

\author[0000-0003-0168-3010]{G.~Mirek Brandt}
\altaffiliation{NSF Graduate Research Fellow}
\affiliation{Department of Physics, University of California, Santa Barbara, Santa Barbara, CA 93106, USA}

\author[0000-0003-4594-4331]{Yunlin Zeng}
\affiliation{School of Physics, Georgia Institute of Technology, 837 State Street, Atlanta, Georgia, USA}

\author[0000-0002-7618-6556]{Daniel Michalik}
\affiliation{European Space Agency (ESA), European Space Research and Technology Centre (ESTEC), Keplerlaan 1, 2201 AZ Noordwijk, The Netherlands}

\author[0000-0001-8170-7072]{Daniella C. Bardalez Gagliuffi}
\affiliation{American Museum of Natural History, 200 Central Park West, New York, NY 10024, USA}

\author[0000-0003-0604-3514]{Virginia Raposo-Pulido}
\affiliation{Space Dynamics Group, Technical University of Madrid, ETSIAE, Pza. Cardenal Cisneros 3, 28040 Madrid, Spain}

\begin{abstract}

We present an open-source Python package, {\it Orbits from Radial Velocity, Absolute, and/or Relative Astrometry} (\codename), to fit Keplerian orbits to any combination of radial velocity, relative astrometry, and absolute astrometry data from the \hipparcos-\gaia Catalog of Accelerations.  By combining these three data types, one can measure precise masses and sometimes orbital parameters even when the observations cover a small fraction of an orbit.
\codename achieves its computational performance with an eccentric anomaly solver five to ten times faster than commonly used approaches, low-level memory management to avoid python overheads, and by analytically marginalizing out parallax, barycenter proper motion, and the instrument-specific radial velocity zero points.  
Through its integration with the \hipparcos and \gaia intermediate astrometry package {\tt htof}, \codename can properly account for the epoch astrometry measurements of \hipparcos and the measurement times and scan angles of individual \gaia\ epochs.  We configure \codename with modifiable {\tt .ini} configuration files tailored to any specific stellar or planetary system. We demonstrate \codename with a case study application to a recently discovered white dwarf/main sequence (WD/MS) system, HD 159062. By adding absolute astrometry to literature RV and relative astrometry data, our comprehensive MCMC analysis improves the precision of HD~159062B's mass by {more than an order} of magnitude to {$0.6083^{+0.0083}_{-0.0073} \Msun$}. We also derive a low eccentricity and large semimajor axis, establishing HD 159062AB as a system that did not experience Roche lobe overflow. 
\end{abstract}


\section{Introduction} \label{sec:intro}

The history of orbit fitting extends back to ancient times, when it culminated in the Ptolemataic model of the Solar system.  During the Age of Enlightenment astronomers fit the first Keplerian orbits to visual binary stars.  More recently, masses derived from orbital fits, known as dynamical masses, anchor models of stellar evolution \citep{Torres+Andersen+Gimenez_2010}.  Precise dynamical masses are used to infer the dynamical evolution of objects that reside off of the main sequence and rapidly evolve in the Hertzsprung-Russell diagram, such as young stars, brown dwarfs, giant planets and white dwarfs \citep{Chabrier_2000,Hillenbrand_2004,Pala_2017}. 

Orbits have assumed a central role in exoplanet research, since exoplanets are often detected only through their effects on their host stars. For a planet detected by radial velocity monitoring, a Keplerian orbital fit may be the only observational result.  Exoplanet demographics show that the mass distribution of companions has a gap between 10 $\Mjup$ and 100 $\Mjup$ \citep{Zucker_2001,Schlaufman_2018}, which suggests that exoplanets form very differently from low-mass stellar companions.
Precise masses, along with luminosities and ages, provide a powerful way to test different planet formation mechanisms, substellar evolutionary models, and white dwarf cooling and atmospheric models \citep[e.g.][]{Benvenuto_1999,Martins_2017}.  {Brown dwarfs and planets cool and fade with time: evolutionary models predict their luminosities as a function of age and mass \citep[e.g.][]{Burrows+Marley+Hubbard+etal_1997,Baraffe+Chabrier+Barman+etal_2003,Saumon+Marley_2008}.  Systems with independent, dynamical masses provide the strongest tests and calibrations of these models \citep{Crepp+Gonzales+Bechter+etal_2016,Dupuy+Liu_2017,Brandt+Dupuy+Bowler_2019,Maire+Baudino+Desidera_2020}.}

For many years, orbital analyses focused on stars and used techniques appropriate for the computing power available. Some approaches used a grid search over a small number of parameters combined with linear least-squares fitting of the remaining parameters to map out $\chi^2$ surfaces \citep[e.g.,][]{1989AJ.....98.1014H,2006AJ....132.2618S}. Other common approaches used non-linear least-squares fitting to all parameters at once \citep[e.g.,][]{1999A&A...351..619F,2001AAS...198.4709G} until Markov Chain Monte Carlo methods became more widespread \citep[e.g.,][]{2007MNRAS.377..415N,2008ApJ...678..463I,2008ApJ...689..436L}.
In the meantime, orbital analysis has become central to the study of exoplanets, inspiring the development of many orbit-fitting tools over the last several years, including $\tt ExoFast$ \citep{Eastman_2013}, $\tt PyAstrOFit$ \citep{Wertz_2016}, {\tt BATMAN} \citep{Kreidberg_2015},
$\tt ExoSOFT$ \citep{Mede_Brandt_2017}, 
$\tt RadVel$ \citep{Fulton_2018} and $\tt orbitize!$ \citep{Blunt_2020}.  
In many cases, only one type of data (astrometry or radial velocity) is available to fit an orbit.  Some of these tools, like {\tt BATMAN} and {\tt RadVel}, are designed with a particular data type in mind, while others including {\tt ExoSOFT} and {\tt orbitize!} are designed to deal with multiple data types.  Multiple types of data can combine to offer much stronger constraints than any one type alone.  Absolute astrometry from \hipparcos and \gaia is now capable of offering meaningful constraints \citep[e.g.,][]{Pearce_2020}, especially when combined with relative astrometry and radial velocity measurements.

The \gaia spacecraft \citep{Gaia_General_2016,Gaia_General_2018,2020GaiaEDR3_catalog_summary} has been surveying the nearby stellar or exoplanetary systems by taking photometric, spectroscopic and astrometric measurements across the sky with on-board instruments since 2014.  The conceptually similar \hipparcos satellite \citep{ESA_1997,vanLeeuwen_2007} obtained astrometric measurements from 1989 to 1993.  \hipparcos and \gaia measured the positions and motions of stars in an inertial reference frame called the International Celestial Reference System (ICRS) defined by distant quasars \citep{Ma_1998,Fey_2015}. The difference in their separate measurements of proper motions indicates acceleration in an inertial frame, which can be used to refine the orbital parameters of the accelerators. However, neither \hipparcos nor \gaia achieved a perfect realization of the ICRS, and both have low-level systematics and uncertainties that are sometimes underestimated.  
\citet{Brandt_2018} has cross-calibrated \hipparcos and \gaia DR2 to account for systematics as a function of position on the sky, putting them in the same reference frame.  {\cite{Brandt_2021} has performed a similar cross-calibration for \gaia EDR3.}

Absolute astrometry can enable precise constraints on systems \citep{Benedict+McArthur+Forveille+etal_2002,Benedict+McArthur+Gatewood+etal_2006} even when only a small fraction of the orbit is observed \citep{Brandt+Dupuy+Bowler_2019}.  The combination of \gaia and \hipparcos proper motions provides a measurement of acceleration in the plane of the sky; a radial velocity trend adds a third dimension to the accelerations. Finally, direct imaging provides projected separations and position angles of the companions, allowing orbital constraints even without observing a substantial fraction of an orbit. 
Long-baseline precision RV surveys can identify stellar, substellar or planetary companions exhibiting accelerations, while 
a growing number of exoplanets are being discovered and characterized via high-contrast imaging \citep{Marois+Macintosh+Barman+etal_2008,Lagrange+Bonnefoy+Chauvin+etal_2010,Rameau+Chauvin+Lagrange+etal_2013,Kuzuhara+Tamura+Kudo+etal_2013,Macintosh+Graham+Barman+etal_2015,Bowler_2016}. 

Many authors have recently combined absolute astrometry with other data types to measure masses and orbits.  These applications include radial velocity-detected exoplanets \citep{Feng+AngladaEscude+Tuomi+etal_2019,Kervella_2020,Damasso+Sozzetti+Lovis+etal_2020,Xuan+Wyatt_2020,DeRosa+Dawson+Nielsen_2020}, directly imaged planets and brown dwarfs \citep{Calissendorff+Janson_2018,Snellen+Brown_2018,Dupuy+Brandt+Kratter+etal_2019,Brandt+Dupuy+Bowler_2019,Nielsen+DeRosa+Wang+etal_2020,Maire+Baudino+Desidera_2020}, and stars \citep{Torres_2019, Torres+Stefanik+Latham_2019,Pearce_2020}. Our goal is to introduce a generalized, optimized, open source and flexible piece of software that can easily incorporate astrometric acceleration measurements.

In this paper, we present \codename, a Python Package designed for fast orbit-fitting and plotting of companions. We adopt a similar approach as \cite{Brandt+Dupuy+Bowler_2019} to jointly fit absolute astrometry, relative astrometry, and/or radial velocities of a given star and companion. Furthermore, we reduce the computational cost of our approach by introducing a more efficient eccentric anomaly solver, marginalizing out four or more nuisance parameters in the likelihood function, and using low-level memory management to avoid python overheads. \codename requires (and is distributed with) the \hipparcos-\gaia Catalog of Accelerations. 

The paper is structured as follows. Section \ref{sec:keplerianorbits} reviews the basics of Keplerian orbits including equations that govern the measured positions and velocities as a function of time and corresponding orbital parameters. This is followed by a discussion of the implementation of these equations in \codename in Section \ref{sec:implementation}. Section \ref{sec:likelihood} describes the marginalized likelihood function over four or more parameters used in MCMC fitting. The computational performance of our implementation is examined in Section \ref{sec:performance}. The configuration and use of \codename, post-processing of the output, and plotting of a suite of eight plots relevant to astrometry and radial velocity are discussed in Section \ref{sec:configuration}. We present a case study application of \codename to a recently discovered WD/MS system HD 159062 in Section \ref{sec:casestudy} where new constraints on the companion mass and the eccentricity are presented. We conclude with our results and findings in Section \ref{sec:conclusion}. 

\section{Keplerian Orbits}
\label{sec:keplerianorbits}

A Keplerian orbit is a solution of the two-body problem for Newtonian gravity.  It is fully described by six orbital elements plus the masses of the two components.  We wish to use measured positions and velocities to derive posterior probability distributions on these eight parameters, marginalizing over several nuisance parameters (e.g., the position, parallax, and velocity of the system's barycenter and astrophysical jitter in the measured radial velocities).  In this section we briefly review the equations that give the measured positions and velocities as a function of time and the orbital parameters.  We will describe our implementation of these equations and of the likelihood in subsequent sections.

In a Keplerian orbit, the mean anomaly $M$ varies linearly with time as 
\begin{equation}
    M = \frac{2\pi}{P} \left(t - t_p \right)
\end{equation}
where $P$ is the system period and $t_p$ is the epoch of periastron passage.  The position and velocity may be computed using the eccentric anomaly, which is given implicitly by
\begin{equation}
    M = E - \varepsilon \sin E
    \label{eq:ecc_anomaly}
\end{equation}
where $\varepsilon$ is the eccentricity.  The radial velocity RV is given through the true anomaly $\nu$ by 
\begin{align}
    \nu &= 2\, {\rm atan2} \left[ \sqrt{1 + \varepsilon} \sin \frac{E}{2}, \sqrt{1 - \varepsilon} \cos \frac{E}{2} \right] \label{eq:trueanomaly} \\
    {\rm RV} &= k \left( \cos \left[ \nu + \omega \right] + \varepsilon \cos \omega \right), \label{eq:rv_naive}
\end{align}
where $\omega$ is the argument of periastron, $k$ is the radial velocity amplitude, and ${\rm atan2}$ is the two-argument arctangent.  We adopt a convention that the orbital parameters all refer to the companion(s).  The orbital parameters for the primary are the same except that $\omega_{\rm pri} = \omega + \pi$.

The projected offset between the two bodies may be computed through the elliptical rectangular coordinates
\begin{align}
    X &= \cos E - \varepsilon \label{eq:ellip_X} \\
    Y &= \left(\sin E \right) \sqrt{1 - \varepsilon^2}, \label{eq:ellip_Y}
\end{align}
and the Thiele-Innes constants
\begin{align}
    A &= \cos \Omega \cos \omega - \sin \Omega \sin \omega \cos i \label{eq:thiele-innes_A} \\
    B &= \sin \Omega \cos \omega + \cos \Omega \sin \omega \cos i \label{eq:thiele-innes_B} \\
    F &= - \cos \Omega \sin \omega - \sin \Omega \cos \omega \cos i \label{eq:thiele-innes_F} \\
    G &= - \sin \Omega \sin \omega + \cos \Omega \cos \omega \cos i. \label{eq:thiele-innes_G}
\end{align}
In Equations \eqref{eq:thiele-innes_A}--\eqref{eq:thiele-innes_G}, $i$ is the inclination, and $\Omega$ is the longitude of the ascending node.  
The projected offsets of the secondary with respect to the primary star in declination $\Delta \delta$ and right ascension $\Delta \alpha* = \Delta (\alpha \cos \delta)$ are then given by 
\begin{align}
    \Delta \delta &= a(AX + FY) \label{eq:offset_Y} \\
    \Delta \alpha* &= a(BX + GY) \label{eq:offset_X}
\end{align}
where $a$ is the semimajor axis in angular units.

Constraints on a system's orbit may come from measurements of the primary star's radial velocity over time, the projected angular offset of the two bodies, and/or the projected motion of either component relative to the system's barycenter in an inertial reference frame.  \codename is designed to account for all of these within the context of Gaussian uncertainties.  While it is straightforward to apply Equations \eqref{eq:rv_naive}, \eqref{eq:offset_Y}, and \eqref{eq:offset_X} to compute predicted positions and velocities, implementing them as shown is typically inefficient and may require many trigonometric function calls per observational epoch.  In the following sections we describe our computational implementation of the equations given in this section and our calculation of the likelihood.

\section{Computational Implementation}
\label{sec:implementation}
\subsection{The Eccentric Anomaly Solver} \label{sec:EAsolve}

Equation \eqref{eq:ecc_anomaly} for the eccentric anomaly is known as Kepler's Equation.  Its importance, combined with its lack of an analytic solution, has inspired centuries worth of work on efficient computational approaches \citep[e.g.][and references therein]{Colwell_1993}.  A first-order Newton-Raphson approach requires an evaluation of sine and cosine at every iteration.  Depending on the quality of the initial guess, this could result in anywhere from a few to $\approx$ten trigonometric evaluations per epoch.  We have developed a more efficient approach for \codename: an eccentric anomaly solver based closely on that of \cite{Raposo-Pulido+Pelaez_2017}, hereafter RPP, with several modifications.  

RPP's basic approach is to obtain a very good initial guess for the eccentric anomaly through fitting formulae, followed by a single step of a modified Newton-Raphson method.  
We begin by reducing the range of the mean anomaly $M$ to $(-\pi, \pi]$; we further reduce the range to $[0, \pi]$ and save the sign of $M$ for later use.  We then adopt the piecewise quintic fitting function for eccentric anomaly $E$ as a function of mean anomaly given in RPP, with the function values and the first two derivatives fixed where the eccentric anomaly is an integer multiple of $\pi/12$ (for twelve polynomials).  Each polynomial $i$ is then defined over a range of mean anomalies $M_i$, $M_{i + 1}$ given in Table 3 of RPP.  To facilitate the calculation of the coefficients, we define the $i^{\rm th}$ quintic polynomial as
\begin{equation}
    E \approx \sum_{k=0}^5 a_{i,k} \left(M - M_i \right)^k
\end{equation}
{where $M_i$ is the minimum mean anomaly in the domain of the $i$-th quintic polynomial.}  With this definition, 
\begin{align}
    a_{i,0} &= E(M_i), \label{eq:a_i0} \\
    a_{i,1} &= E^\prime(M_i),~~{\rm and} \label{eq:a_i1} \\
    a_{i,2} &= \frac{1}{2} E^{\prime\prime}(M_i), \label{eq:a_i2}
\end{align}
where the values and derivatives are tabulated in Table 3 of RPP.  The other three coefficients may be derived by matching the values at $M_{i + 1}$,
\begin{align}
    E(M_{i + 1}) &= \sum_{k=0}^5 a_{i,k} \left(M_{i + 1} - M_i \right)^k, \label{eq:a_i3} \\
    E^\prime(M_{i + 1}) &= \sum_{k=1}^5 a_{i,k} k \left(M_{i + 1} - M_i \right)^{k - 1},~~{\rm and}  \label{eq:a_i4} \\
    E^{\prime\prime}(M_{i + 1}) &= \sum_{k=2}^5 a_{i,k} k(k - 1)\left(M_{i + 1} - M_i \right)^{k - 2}. \label{eq:a_i5} 
\end{align}
Since $a_{i,0}$, $a_{i,1}$, and $a_{i,2}$ are known from Equations \eqref{eq:a_i0}--\eqref{eq:a_i2}, Equations \eqref{eq:a_i3}--\eqref{eq:a_i5} form a linear system of three equations for the three remaining coefficients.  We solve this linear system by hand.

With a good initial guess for the eccentric anomaly $E$, we deviate very slightly from RPP and use Halley's method to refine the solution for eccentricities $\varepsilon < 0.78$.  This requires the sine and cosine of the eccentric anomaly.  We use a series expansion for one and a square root call for the other; these are significantly faster than trigonometric calls and {remain within a factor of a few of double precision floating point accuracy.}
If $E < \pi/4$, we use the Taylor series for sine up to $15^{\rm th}$ order for a cost of seven additions and nine multiplications.  We then compute $\cos E = \sqrt{1 - \sin^2 E}$, with no ambiguity in the sign given the range reduction.  If $E > 3\pi/4$, we use $\pi - E$ in the same Taylor expansion for sine.  If $\pi/4 < E < 3\pi/4$, we use the identity $\cos E = \sin (\pi/2-E)$ and the Taylor expansion of sine to compute $\cos E$; we then use a square root call to evaluate $\sin E$.  

For $\varepsilon < 0.78$, a single iteration of Halley's method brings the accuracy of our computed eccentric anomaly to nearly double precision.  We then update $\sin E$ and $\cos E$, {which we initially computed using series and square roots}, with the summation trigonometric identities for $\sin (E + \delta E)$ and $\cos (E + \delta E)$. {We evaluate} $\sin \delta E$ and $\cos \delta E$ to second order in keeping with the {second order} accuracy of {Halley's method}.  {Finally, we multiply $E$ and $\sin E$ by the sign of the mean anomaly.}  This brings $E$, $\sin E$, and $\cos E$ all close to double precision, with worst case errors below $10^{-15}$ for $\varepsilon < 0.78$.

\begin{figure*}
    \includegraphics[width=0.5\linewidth]{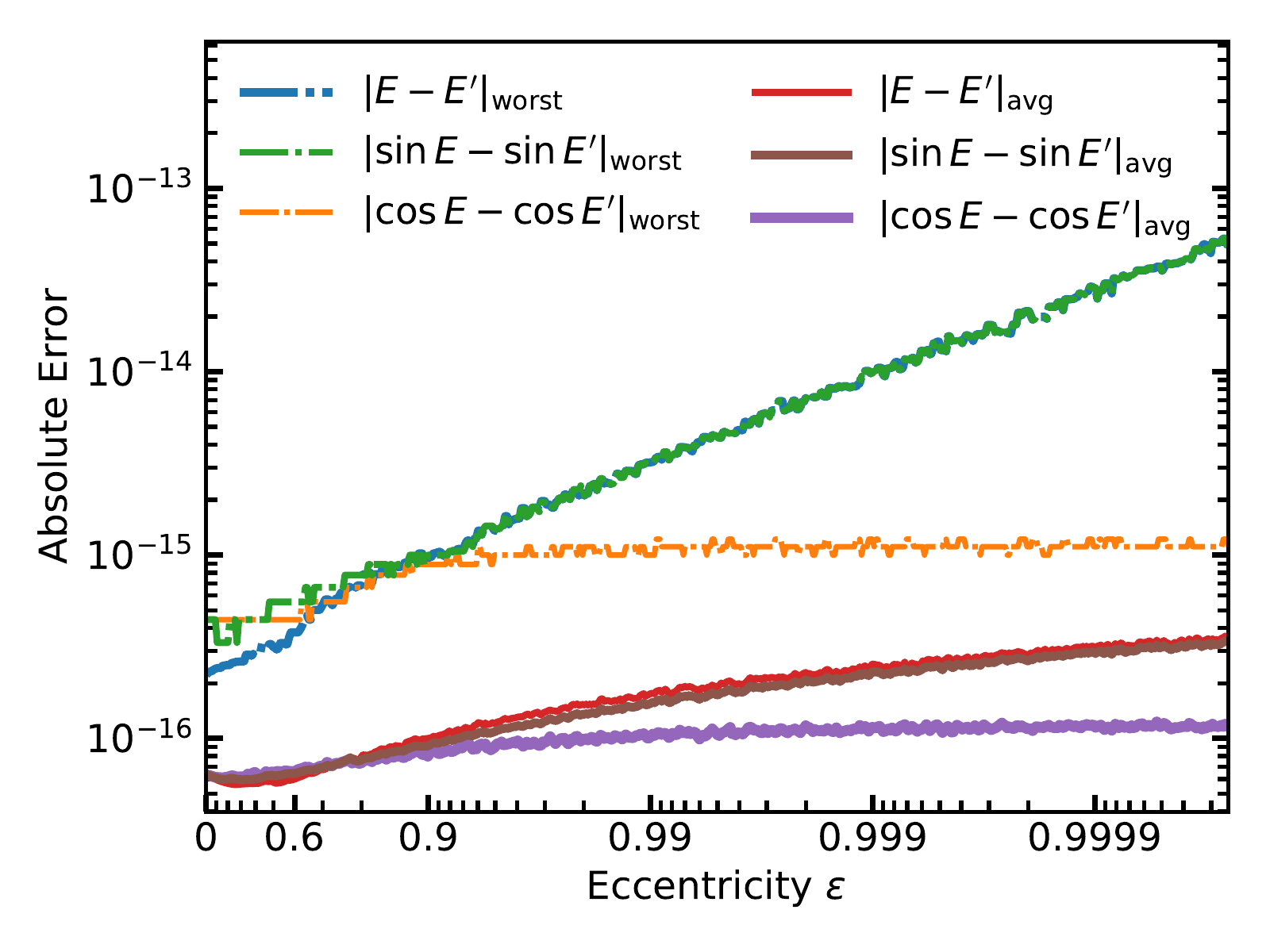}
    \includegraphics[width=0.5\linewidth]{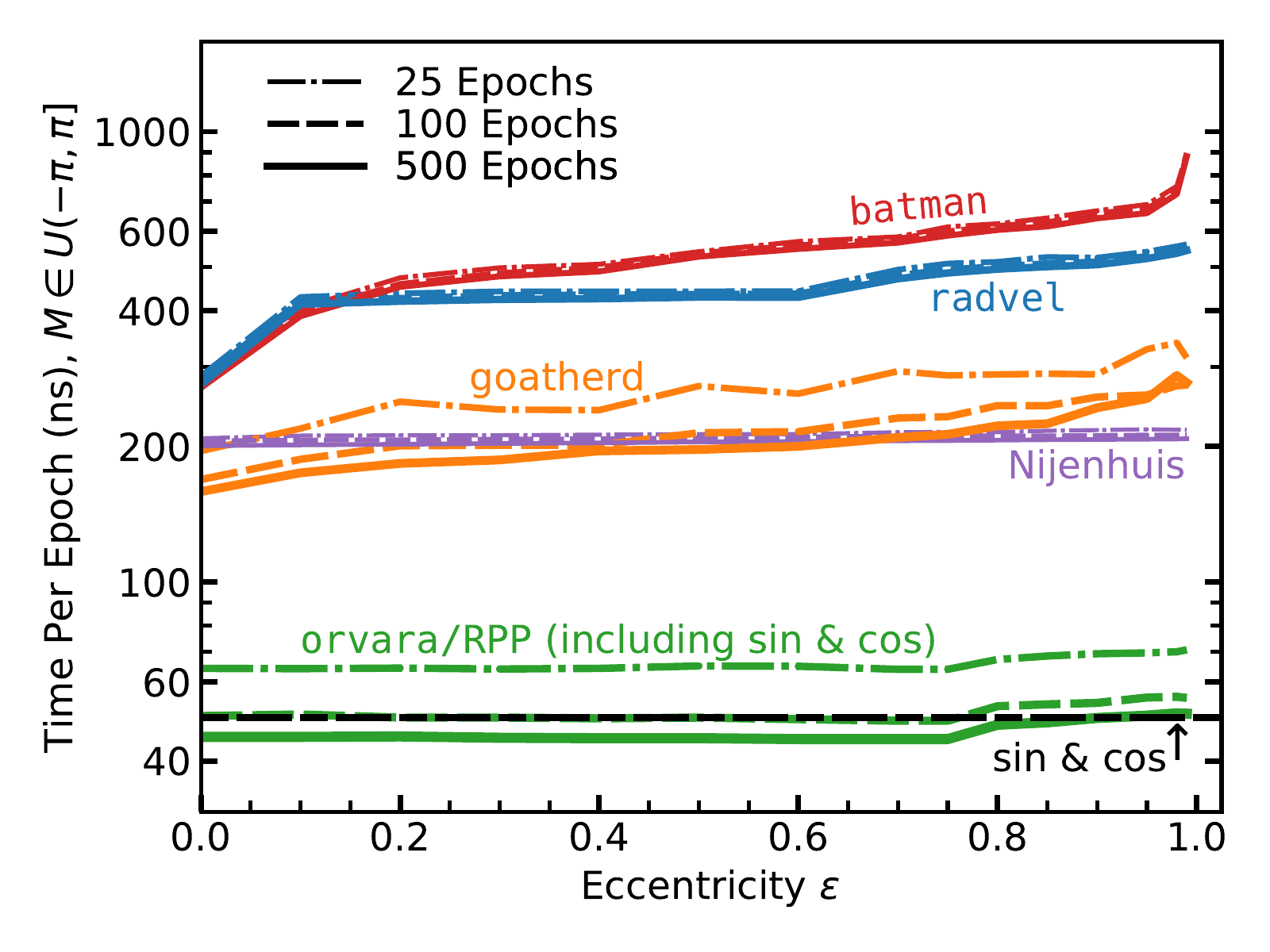}
    \caption{Left: residuals of the eccentric anomaly in radians, its sine and cosine, as a function of eccentricity for a uniform distribution of mean anomalies.  The increasing errors in $E$ and $\sin E$ as $\varepsilon$ approaches unity are due to their mutual cancellation in Kepler's equation: the residuals in $E - \varepsilon \sin E$ remain below $10^{-15}$.  Right: time per epoch for the eccentric anomaly solvers in \codename, {\tt radvel} \citep{Fulton_2018}, {the \cite{Nijenhuis_1991} solver of {\tt exoplanet} \citep{2019zndo...1998447F}, the goatherd/contour integral approach of \cite{Philcox+Goodman+Slepian_2021},} and {\tt batman} \citep{Kreidberg_2015} using a single core of an Intel Xeon E5-2630, 2.2~GHz processor, and for 25, 100, or 500 epochs per eccentricity (dot-dashed, dashed, and solid lines, respectively).  Our implementation of the \cite{Raposo-Pulido+Pelaez_2017} algorithm includes the calculation of sine and cosine; we add a calculation of sine and cosine to the other methods.  
    The dashed black curve shows the computational cost of a single call each to sine and cosine for each epoch.  Our RPP implementation also has an additional overhead of $\sim$400~ns/orbit for the allocation and computation of the polynomials for the initial guess.  {This, combined with small Python overheads, accounts for the difference in cost between 25, 100, and 500 epochs.}}
    \label{fig:EAsolver}
\end{figure*}

At higher eccentricities, the initial polynomial guess behaves poorly when $M \ll 1$.  If $2M + (1 - \varepsilon) < 0.2$, we adopt the second-order series expansion derived in RPP (their Equation (19)) for our starting point, and revert to the quintic polynomial fits otherwise.  We find that at higher eccentricities and low mean anomalies, we sometimes require a third-order step to reach double precision accuracy in a single iteration.  We {therefore} use Householder's third-order method if both $\varepsilon > 0.78$ and $M < 0.4$, and Halley's second-order method otherwise.  As before, we refine $\sin E$ and $\cos E$ using the summation identities.  Where {$\varepsilon > 0.78$ and $M < 0.4$}, we use series expansions up to third order to match the overall accuracy of Householder's {(third order)} method.  This approach is slightly slower than at lower eccentricities due to the higher-order variant of Newton's method and to the need to occasionally use a series expansion for the initial guess.  Up to eccentricities of 0.9999, however, the absolute errors in $E$ and $\sin E$ remain below $2 \times 10^{-14}$ at all mean anomalies, and up to eccentricities of 0.99, they remain below $3 \times 10^{-15}$ for all mean anomalies.

Figure \ref{fig:EAsolver} summarizes the performance of the eccentric anomaly solver.  The left panel shows the overall accuracy, both the average and maximum errors as a function of eccentricity for a uniform distribution of eccentric anomalies.  Even at an eccentricity of 0.9999 the algorithm is accurate to $2 \times 10^{-14}$ in $E$ and $\sin E$, and to $10^{-15}$ in $\cos E$.  The increasing error at high eccentricity is a result of a cancellation between $E$ and $\sin E$ in Kepler's equation; it could be overcome by rewriting the equation in terms of $1 - \varepsilon$ and expanding $\sin E$ as a series for small values of $E$.  However, this would not solve the more fundamental problem that we can only compute the mean anomaly from the observation epoch with limited precision, and that $dE/dM$ can be large for $\varepsilon \approx 1$.

The right panel of Figure \ref{fig:EAsolver} shows the computational cost of our implementation of RPP compared to some other widely-used eccentric anomaly solvers, all evaluated on a single core of an Intel Xeon E5-2630, 2.2~GHz processor.  Our implementation gives $\sin E$ and $\cos E$ as byproducts of the calculation.  We add a call to sine and cosine when assessing other methods, as $\sin E$ and $\cos E$ are required to calculate radial velocities and/or positional offsets.  
There is an overhead of $\approx$170~ns associated with calculating the 72 polynomial coefficients at a given eccentricity, making the performance of our RPP implementation only slightly faster than other solvers for a single epoch.  We have also implemented a version of this solver tailored to single epoch calculations.  The single epoch version eliminates most of the overhead by constructing only one of the quintic polynomials (six coefficients).  All methods shown in the right panel of Figure \ref{fig:EAsolver} also incur an overhead of a few hundred ns for the Cython function call.  For 25 epochs at the same eccentricity, our implementation of RPP is around five times faster than these other approaches.  With $\gtrsim$100 epochs at a given eccentricity, it can be faster than one call per epoch to both sine and cosine.  Our implementation is $\sim$5--8 times faster than the SDG code \citep{Raposo-Pulido+Pelaez_2017} because of its avoidance of explicit trigonometric calls and its approach to computing the polynomial coefficients.

\subsection{Radial Velocities}

Given the eccentric anomaly $E$, eccentricity $\varepsilon$, and radial velocity amplitude $k$, the radial velocity can be calculated by Equations \eqref{eq:trueanomaly} and \eqref{eq:rv_naive}.
Written this way, the radial velocity evaluation requires five trigonometric calls (four if range reduction is used to enable the use of the single-argument arctangent).  Only one of these, $\cos \omega$, can be computed once for all epochs in an orbit.  However, Equations \eqref{eq:trueanomaly} and \eqref{eq:rv_naive} are equivalent by trigonometric identities to
\begin{align}
    g &= \left(\sqrt{\frac{1 + \varepsilon}{1 - \varepsilon}}\right) \left(\frac{1 - \cos E}{\sin E}\right) \label{eq:rv_g} \\
    {\rm RV} &= k \left( \left(\frac{1 - g^2}{1 + g^2} + \varepsilon \right) \cos \omega - \left(\frac{2g}{1 + g^2} \right) \sin \omega \right) \label{eq:rv_new} .
\end{align}
The quantities $\sin \omega$ and $\cos \omega$ appearing in Equation \eqref{eq:rv_new} only need to be computed once per {\it orbit}, not once per epoch.  The quantities $\sin E$ and $\cos E$, needed for Equation \eqref{eq:rv_g},
were already evaluated in the course of solving for the eccentric anomaly.  Once $\sin \omega$ and $\cos \omega$ are computed for an orbit, computing the radial velocities using Equations \eqref{eq:rv_g} and \eqref{eq:rv_new} does not require a single trigonometric evaluation.  In comparison tests, we found the use of Equations \eqref{eq:rv_g} and \eqref{eq:rv_new} to be $\sim$10--20 times faster than Equations \eqref{eq:trueanomaly} and \eqref{eq:rv_naive}.

Equation \eqref{eq:rv_g} can be problematic when $E \approx 0$.  If $|E| < 0.015$, we use the series expansion up to fifth order (there is no sixth order term); this gives a value of $g$ accurate to better than $10^{-15}$.  If $E \approx \pm \pi$, Equation \eqref{eq:rv_new} presents no numerical problems unless $\sin E$ is precisely zero (in which case we simply set ${\rm RV} = k (\varepsilon - 1) \cos \omega$).

\subsection{Absolute and Relative Projected Positions}
\label{subsec:absolute_relative_astrometry}

The relative offsets of the secondary from the primary star in right ascension $\alpha*$ and declination $\delta$ (where $\alpha* = \alpha \cos \delta$) are given by Equations \eqref{eq:offset_Y} and \eqref{eq:offset_X}.  These equations present no computational difficulties.  With $\sin E$ and $\cos E$ already computed from the eccentric anomaly solver, Equations \eqref{eq:ellip_X}--\eqref{eq:offset_X} require five trigonometric evaluations plus one square root per orbit; zero per epoch.  

We typically fit orbits using both absolute and relative astrometry.  The displacement of the primary star from the system's barycenter is related to the relative separations of Equations \eqref{eq:offset_Y} and \eqref{eq:offset_X} by
\begin{align}
    \Delta \delta_{\star} &=  \left( \frac{-M_{\rm B}}{M_{\rm A} + M_{\rm B}} \right) \Delta \delta \label{eq:absDec} \\
    \Delta \alpha*_{\star} &=  \left( \frac{-M_{\rm B}}{M_{\rm A} + M_{\rm B}} \right) \Delta \alpha* \label{eq:absRA}
\end{align}
where $M_{\rm A}$ is the mass of the primary star and $M_{\rm B}$ is the mass of its companion. 

Astrometric missions like \hipparcos and \gaia measure the position of a star many times and fit an astrometric sky path.  We use the Hundred Thousand Orbit Fitter ({\tt htof}) \citep{Brandt+Michalik+Brandt+etal_2021} to compute synthetic \hipparcos and \gaia catalog positions and proper motions from 
the offsets given in Equations \eqref{eq:absDec} and \eqref{eq:absRA}. We then compare the {\tt htof} synthetic catalog values to the cross-calibrated absolute astrometry of HGCA \citep{Brandt_2018}.  We refer the reader to the source code at \url{https://github.com/gmbrandt/htof} and the publication (when available) for further details.

\subsection{Multiple Companions} \label{subsec:multiple_companions}

Keplerian orbits only describe two-body systems.  To fit more than one companion in \codename, we use a set of simplifying approximations that we describe here. 

We approximate the star's motion by a superposition of Keplerian orbits, one due to each companion.  We use the orbital elements of each companion (shifting the argument of periastron by $\pi$ to give the stellar orbit), but we modify the total mass.  For the mutual orbit between the star and a given companion indexed by $i$, we add the mass of all other companions that orbit closer to the star than companion $i$ itself.  Effectively, we approximate the star and inner companions as a single body and solve for its motion about its barycenter due to the $i$-th companion.  We then solve for the orbital motion within the inner system when we account for those companions.  We add the contributions to the host star's motion from all companions for both absolute astrometry and radial velocity.

We treat relative astrometry slightly differently.  In this case, we ignore the influence of all companions orbiting beyond the companion of interest.  We first compute the relative astrometry between this companion and the barycenter of the star and all inner companions.  We then add the offset of the star relative to this barycenter due to the inner companions.  This calculation is the same as the shift in absolute astrometry due to these inner companions.  

These approximations do not fully capture interactions between companions.  
While incorrect in detail, our approach does capture some of the effects of a multi-body system.  It can be used to fit extremely accurate relative astrometry (like that from GRAVITY, \citealt{Gravity+Nowak+Lacour+etal_2020}), though not with the fidelity of a full (and expensive) suite of $N$-body simulations.

\section{The Likelihood}
\label{sec:likelihood}

We compute the likelihood ${\cal L}$ of an orbit as 
\begin{align}
    -2 \ln {\cal L} = \chi^2 = \chi^2_{\rm RV} + \chi^2_{\rm rel\,ast} + \chi^2_{\rm abs\,ast} .
\end{align}
We treat $\chi^2_{\rm RV}$, the radial velocity component, first.  We then take the latter two terms together as we marginalize out the barycenter's proper motion and the system's parallax.

\subsection{Radial Velocity}

For the radial velocity, we take
\begin{align}
    \chi^2_{\rm RV} = \sum_{j=1}^{N_{\rm inst}}\sum_{k=1}^{N_{\rm RV}} \bigg(& \frac{\left({\rm RV}_{k}+{\rm ZP}_j-{\rm RV}\left[t_k\right] \right)^2}{\sigma^2[{\rm RV}_k] + \sigma^2_{\rm jit}} \label{eq:chisq_RV}  \nonumber \\
    &\quad + \ln \left[ \sigma^2[{\rm RV}_k] + \sigma^2_{\rm jit} \right] \bigg),
\end{align}
where ${\rm ZP}_j$ is the instrument-specific radial velocity zero point and $\sigma^2_{\rm jit}$ is a jitter term.  {We use ${\rm RV}_k$ to denote the measured RV at epoch $t_k$, ${\rm RV}[t_k]$ for the model-predicted RV and $\sigma^2[{\rm RV}_k]$ for its variance.}

{Our approach} differs from \cite{Brandt+Dupuy+Bowler_2019} in that we adopt a single jitter for all instruments, attributing it to stellar activity.  This has the disadvantage of assuming that instrumental uncertainties are properly estimated and that instruments sample similar regimes of stellar activity (the latter assumption could be problematic if combining visible and near-infrared data).  However, it makes the fit robust to the inclusion of instruments with just a few data points, and it reduces the number of parameters to fit.  {\codename does include the option of fitting a different jitter to every RV instrument. }

Equation \eqref{eq:chisq_RV} is the only place in the likelihood where the radial velocity zero point appears.  We therefore marginalize it out, assuming a flat prior, by integrating
\begin{equation}
    {\cal L} \propto \int_{-\infty}^\infty d {\rm ZP} \exp \left[ -\frac{\chi^2_{\rm RV}}{2} \right].
\end{equation}
We perform the integral separately for each instrument.  This requires only one pass through the radial velocity data set, and adds a negligible amount of computation.  Performing the integral results in replacing $\chi^2_{\rm RV}$ with
\begin{align}
    \chi^2_{\rm RV} = \sum_{j=1}^{N_{\rm inst}} &\left( -\frac{B_j^2}{4A_j} + C_j + \ln A_j \right) \nonumber \\
    &+ \sum_{j=1}^{N_{\rm inst}}\sum_{k=1}^{N_{\rm RV}} \ln \left[  \sigma^2[{\rm RV}_k] + \sigma^2_{\rm jit} \right] \label{eq:chisq_RV_marginalized}
\end{align}
where
\begin{align}
    A_j &= \sum_{k=1}^{N_{\rm RV}} \frac{1}{\sigma^2[{\rm RV}_k] + \sigma^2_{\rm jit}}, \\
    B_j &= \sum_{k=1}^{N_{\rm RV}} \frac{2 \left({\rm RV}_{k} - {\rm RV}\left[t_k\right] \right)}{\sigma^2[{\rm RV}_k] + \sigma^2_{\rm jit}}, \\
    C_j &= \sum_{k=1}^{N_{\rm RV}} \frac{\left({\rm RV}_{k} - {\rm RV}\left[t_k\right] \right)^2}{\sigma^2[{\rm RV}_k] + \sigma^2_{\rm jit}} \label{eq:rv_c},
\end{align}
and the sums over $k$ are restricted to radial velocity data points from instrument $j$.  
\codename uses Equations \eqref{eq:chisq_RV_marginalized}--\eqref{eq:rv_c} as shown.

\subsection{Astrometry}
\label{subsec:astrometry}
The $\chi^2$ for relative astrometry consists of two components: the relative separation $\rho$ and the position angle $\theta$ measured east of north.  The model orbit's relative separation is the product of the projected relative separation $\rho$ in AU and the system's parallax $\varpi$.  In general, the measurements of separation and position angle may be covariant: we take $c_{\rho \theta,k} \in (-1, 1)$ to be the correlation coefficient between the two measurements at epoch $k$.  The contribution to $\chi^2$ is then
\begin{align}
    \chi^2_{\rm rel\,ast} &=  \sum_{k=1}^{N_{\rm ast}} \frac{\lfloor\theta_k - \theta[t_k]\rfloor^2}{(1 - c^2_{\rho \theta,k}) \sigma^2[\theta_k]} + \sum_{k=1}^{N_{\rm ast}} \frac{\left( \rho_k-\varpi\rho\left[t_k\right] \right)^2}{(1 - c^2_{\rho \theta,k})\sigma^2 [\rho_k]} \nonumber \\
    &\qquad -2 \sum_{k=1}^{N_{\rm ast}}\frac{c_{\rho \theta,k} \lfloor\theta_k - \theta[t_k]\rfloor\left( \rho_k-\varpi\rho\left[t_k\right] \right)}{(1 - c^2_{\rho \theta,k})\sigma[\theta_k]\sigma[\rho_k]}.
    \label{eq:chisq_relast}
\end{align}
where {$\rho_k$ and $\theta_k$ are the observed separation and position angle at time $t_k$, $\rho[t_k]$ and $\theta[t_k]$ are the model-predicted values, and} $\lfloor \theta_k - \theta[t_k] \rfloor$ is the difference between the measured and the predicted position angles, reduced to the range $(-\pi,\pi]$.

The absolute astrometry in angular units is similarly proportional to parallax, and also has a velocity zero point.  This is the proper motion of the system barycenter in the plane of the sky $\overline{\bm \mu}$, an almost perfect analog of the radial velocity zero point.  This component of the likelihood then reads 
\begin{align}
    \chi^2_{HG} = &\left( {\bm \mu_{H, \rm o}} - \overline{\bm \mu} - \varpi \bm{\mu}_{H} \right)^T {\bf C}_H^{-1} \left( {\bm \mu_{H, \rm o}} - \overline{\bm \mu} - \varpi \bm{\mu}_{H} \right) \nonumber \\
    &+\left( {\bm \mu_{HG, \rm o}} - \overline{\bm \mu} - \varpi \bm{\mu}_{HG} \right)^T {\bf C}_{HG}^{-1} \left( {\bm \mu_{HG, \rm o}} - \overline{\bm \mu} - \varpi \bm{\mu}_{HG} \right) \nonumber \\
    &+\left( {\bm \mu_{G, \rm o}} - \overline{\bm \mu} - \varpi \bm{\mu}_{G} \right)^T {\bf C}_G^{-1} \left( {\bm \mu_{G, \rm o}} - \overline{\bm \mu} - \varpi \bm{\mu}_{G} \right)
    \label{eq:chisq_absast}
\end{align}
where, e.g., $\bm{\mu}_{H,\rm o}$ is the observed \hipparcos proper motion and $\bm{\mu}_{H}$ is the model orbit's predicted \hipparcos proper motion in AU\,yr$^{-1}$ (which must then be multiplied by the parallax to obtain a proper motion in angular units).  

In some cases, \gaia obtains an astrometric solution for both stars in a binary.  In this case, the observed epochs and scan angles will generally be the same for both stars, and the position and proper motion of the secondary may be easily computed from those of the primary.  In the frame of the system barycenter (in which the orbit code operates), we have 
\begin{equation}
    \bm{\mu}_{\rm B} = -\bm{\mu} \left( \frac{M_{\rm A}}{M_{\rm B}} \right)
\end{equation}
where $\bm{\mu}$ is the proper motion of the primary star and $\bm{\mu}_{\rm B}$ is the proper motion of its companion.  Denoting the \gaia proper motion of the secondary by ${\bm \mu_{G, \rm o, B}}$, this measurement contributes an extra term to the astrometric $\chi^2$,
\begin{equation}
    \Delta \chi^2_{HG} = \left( {\bm \mu_{G, \rm o, B}} - \overline{\bm \mu} - \varpi \bm{\mu}_{G, \rm B} \right)^T {\bf C}_{G, \rm B}^{-1} \left( {\bm \mu_{G, \rm o, B}} - \overline{\bm \mu} - \varpi \bm{\mu}_{G, \rm B} \right)
    \label{eq:dchisq_secondary}
\end{equation}
where ${\bf C}_{G, \rm B}^{-1}$ is the inverse of the \gaia covariance matrix for the secondary and $\overline{\bm \mu}$ remains the center-of-mass motion of the system's barycenter.  Including this additional term assumes that both stars have their proper motion in the same reference frame and have well-calibrated errors.  Either assumption may break down in practice: \gaia DR2 has magnitude-dependent systematics in the reference frame of $\sim$0.2~mas\,yr$^{-1}$ \citep[Figure~4 of][]{Gaia_Astrometry_2018}, while the HGCA required spatially variable error inflation by a typical factor $\sim$1.7 \citep{Brandt_2018}.  {\gaia EDR3 retains magnitude-dependent systematics of up to $\approx$100\,$\mu$as\,yr$^{-1}$ \citep{Cantat-Gaudin+Brandt_2021}.} We also neglect covariance between the secondary's measured proper motion and the projected separation of the two bodies.

Equations \eqref{eq:chisq_relast} and \eqref{eq:chisq_absast} (with or without the additional term from Equation \eqref{eq:dchisq_secondary}) are quadratic equations in parallax and the two components of the proper motion of the system's barycenter.  We adopt a uniform prior on $\bm{\overline\mu}$, but use a Gaussian prior in parallax to incorporate the measured \gaia value.  This is equivalent to adding one additional component to the log likelihood,
\begin{equation}
    \chi^2_{\varpi} = \frac{\left( \varpi - \varpi_{\it Gaia} \right)^2}{\sigma_{\varpi,\it Gaia}^2}. \label{eq:parallax_prior}
\end{equation}
With the addition of Equation \eqref{eq:parallax_prior}, we can write the design matrix of the system and solve for the maximum likelihood values of the parameters and their covariance matrix.  Integrating the likelihood over parallax and barycentric proper motion is equivalent to substituting these maximum likelihood values into the expressions for $\chi^2$ and multiplying the likelihood by the square root of the determinant of the covariance matrix.  This requires solving a $3 \times 3$ linear system and computing the determinant of a $3 \times 3$ matrix, neither of which incurs a significant computational cost.  We write out this linear system in the Appendix.  We solve the $3 \times 3$ linear system using the singular value decomposition, but with a C routine to avoid the overheads associated with python function calls and memory management.

\subsection{The Marginalized Likelihood}

The integrations discussed in this section remove four parameters from the fit for a system when all radial velocities are from a single instrument, and more parameters otherwise.  The computational cost of doing so is small relative to the cost of computing the orbit.  The maximum posterior probability values of the parallax and barycenter proper motion may be of interest and are difficult to reconstruct from an output chain.  We therefore save these quantities at each step for later use, along with the values for the \hipparcos, {\it Hipparcos--Gaia}, \gaia, relative separation, position angle components of $\chi^2$, and maximum likelihood radial velocity zero points of all instruments.

It is much more difficult to analytically integrate out other parameters.  \cite{Wright+Howard_2009} show that the log likelihood can be linear in additional parameters if the parallax is accurately known.  However, the priors on the combinations of parameters that enter the problem linearly can be complex even for simple choices of the underlying Keplerian parameters; the product of the likelihood and prior is difficult to integrate.  Introducing relative astrometry (as we do for many of our targets) breaks the linearity entirely.  If we consider parallax to be known, we may still integrate out the secondary mass if we adopt the total system mass as the other mass parameter in the problem, as long as we use a flat prior.  Given these restrictions, we retain parallax as a linear parameter and do not analytically integrate out any others.

Our likelihood, after marginalizing four parameters, is a function of the six Keplerian orbital elements, the masses of the two components, and a radial velocity jitter, for a total of nine parameters.  If there is more than one companion in the system, the likelihood becomes a function of two parameters (mass of the primary and radial velocity jitter) plus seven per companion (the six Keplerian orbital elements and the companion's mass).  We sample from this marginalized likelihood using the parallel-tempering MCMC sampler {\tt ptemcee} \citep{Vousden+Farr+Mandel_2016}, a fork of {\tt emcee} \citep{Foreman-Mackey+Hogg+Lang+etal_2013}.  

\section{Performance}
\label{sec:performance}
The computational cost of a single step includes contributions from the computation of the eccentric anomaly, the radial velocity and positional offsets, and the likelihood calculation.  A typical data set consists of a few hundred measurements.  Of these, $\sim$150--200 might be individual epochs from \hipparcos and \gaia.  For a sample data set of Gl~758 \citep{Brandt+Dupuy+Bowler_2019}, there are 652 radial velocity measurements, 181 epochs for the absolute astrometry, and 4 epochs for relative astrometry.  This data set requires 85~$\mu$s per iteration on a single core of a 2.2~GHz Intel Xeon E5-2630.  Of this time, 44~$\mu$s is spent in the calculation of the eccentric anomalies, 5~$\mu$s on radial velocities, 2~$\mu$s on absolute astrometry for individual epochs, 9~$\mu$s on fitting the \hipparcos and \gaia models to the epoch astrometry, and 10~$\mu$s on the likelihood calculation.  The remaining 15~$\mu$s are in overheads associated with \texttt{emcee} and the main python functions.  For a sample data set of HD~4747, with 49 radial velocity measurements, 159 epochs for the absolute astrometry, and 8 epochs for relative astrometry, the total cost is $\sim$40~$\mu$s per iteration.  Due to the smaller number of data points, the times spent calculating the eccentric anomaly, radial velocity, and likelihood fall to 11~$\mu$s, 1~$\mu$s, and 5~$\mu$s, respectively.  The cost of fitting the \hipparcos and \gaia models to the epoch astrometry remains $\sim$9~$\mu$s.  By comparison, a single-planet fit to 216 epochs of radial velocity data using \texttt{radvel} takes 1.0~ms per iteration with one core of the same CPU, or about 25 times slower than a fit to the same number of epochs (but with a diversity of data types) using \codename. 

Overheads associated with allocating \texttt{numpy} arrays could take the majority of the computational time if we operated exclusively on arrays.  We therefore explicitly handle memory allocation in Cython, and use C structures with pointers rather than \texttt{numpy} arrays.  Replacing each pointer allocation with an \texttt{ndarray} and creating a memoryview would more than double the run time for HD~4747.  For similar reasons, we use a C implementation of the singular value decomposition rather than a python call to the linear algebra routine in \texttt{numpy}.  For a typical system with a few hundred observational epochs, the total computational cost to compute an orbit and evaluate its likelihood (marginalizing out parallax, radial velocity zero point, and barycenter proper motion) is equivalent to about three to seven trigonometric evaluations per epoch.

\codename uses the parallelization built into the packages \texttt{emcee} and {\tt ptemcee}.  The scaling under typical use is poor when using more than a handful of processors. {Using five processors results in a speedup by just a factor of two on our machines, while adding further processors gives almost no additional speedup at all.  Parallelization also works differently on different operating systems, and may not be supported on all machines.  Regardless,} multiple instances of the program may be run simultaneously for significantly better performance.

\section{Configuration, Use and Plotting}
\label{sec:configuration}

\codename is an open-source Python package that performs parameter fits for multi-planetary or binary star systems using a combination of the \hipparcos-\gaia Catalog of Accelerations, literature radial velocities, and/or relative astrometry. \codename includes Python plotting routines to produce a comprehensive orbit-fitting and plotting package. It is available on GitHub\footnote{https://github.com/t-brandt/orvara}. The detailed installation procedure of \codename and its updated documentation can be found in the $\tt README.rst$ file. 

{Most applications of \codename use radial velocity and/or relative astrometry data.  Our example in Section \ref{sec:casestudy} uses both.  Each {type of data} is given in an input ASCII text file with fields separated by spaces or tabs; lines beginning with {\tt \#} are ignored.  We provide examples in the repository and summarize their required structure in \autoref{tab:fileformats}.  The radial velocity file must include BJD, RV, and RV error (both in units of m/s) as its first three columns.  An optional fourth column may include an integer $\geq$0 to distinguish instruments from one another; each instrument will have its own unique RV zero point.  If the fourth column is not supplied \codename will assume that all radial velocities share the same zero point.  The relative astrometry file must include date (either BJD or decimal year, with values $<$3000 interpreted as years), separation and its error (both in arcseconds), and position angle and its error (east of north, both in degrees) as its first five columns.  An optional sixth column gives the correlation coefficient ($\in (-1, 1)$) between separation and position angle.  An optional seventh column gives the ID of the measured companion.  The first companion is indexed as 0, the second as 1, etc.  If the sixth and seventh columns are not supplied, \codename will use default values of 0 for each.}

\begin{deluxetable*}{lcccr}
\tablecaption{Format of the input data files \label{tab:fileformats}}
\tablewidth{0pt}
\tablehead{Column Number
& Description
& Units
& Required?
& Default Value
}
\startdata
\multicolumn{5}{c}{Radial Velocity Data File} \\ \hline
1 & Observation epoch & BJD & yes & \ldots \\
2 & Radial velocity & m/s & yes & \ldots \\
3 & Radial velocity error & m/s & yes & \ldots \\
4 & RV instrument ID\tablenotemark{a} & \ldots & no & 0 \\ \hline
\multicolumn{5}{c}{Relative Astrometry Data File} \\ \hline
1 & Observation epoch & Decimal year or BJD & yes & \ldots \\
2 & Angular Separation & arcsec & yes & \ldots \\
3 & Separation error & arcsec & yes & \ldots \\
4 & Position angle (E of N) & degrees & yes & \ldots \\
5 & Position angle error & degrees & yes & \ldots \\
6 & Sep/PA correlation coefficient\tablenotemark{a} & \ldots & no & 0 \\
7 & Companion ID\tablenotemark{a} & \ldots & no\tablenotemark{b} & 0 
\enddata
\tablenotetext{a}{Valid values: sequential integers from 0 to $n_{\rm inst} - 1$ for RV instrument ID, integers from 0 to $n_{\rm companions} - 1$ for companion ID, real numbers between $-1$ and 1 for sep/PA correlation coefficient.}
\tablenotetext{b}{Column 6 must be included (even if all zeros) when using column 7 for companion ID.}
\end{deluxetable*}

Orbit fitting and plotting can be accessed with the $\tt fit\_orbit$ and $\tt plot\_orbit$ commands from the command line, respectively. The $\tt fit\_orbit$ output is a single Flexible Image Transport System (FITS) file \citep{Wells+Greisen+Harten_1981} containing the MCMC chains and described in more detail below. The header of this FITS file includes the configuration parameters used to produce the chain.  The $\tt plot\_orbit$ output is a suite of up to eight plots relevant to RV and relative astrometry and also discussed individually below. A star-specific {\tt .ini} configuration file containing the appropriate file directories and settings serves as the input for both commands. This configuration file is comprised of four main sections denoted in square brackets: the data paths to all related file directories ($\tt [data\_paths]$), settings for MCMC fitting ($[\tt mcmc\_settings]$), settings for the primary mass prior and uncertainty ($\tt [priors\_settings]$), settings for plotting ($\tt [plotting]$), and for saving a table of results ($\tt save\_results$). A complete list of file directories and customizable input parameters including their descriptions and functionalities is provided in \autoref{tab:configfile}. Sample configuration files for Gl 758 and HD 4747, along with their sample data sets of radial velocity and relative astrometry, are included in the package.  {A user can omit data (for example, if there is no relative astrometry) by providing a blank or invalid file path.  A user can also omit absolute astrometry by providing an invalid \hipparcos ID and supplying an explicit prior on the system parallax.}

The first few lines in the {\tt .ini} file specify the locations of data files and the \hipparcos ID of the primary star.  The initial guesses for the nine MCMC parameters (plus seven for each companion beyond the first) may also be set in a file referenced in the {\tt .ini} configuration file.  \codename draws starting values for each walker from a normal or lognormal distribution with the means and variances specified.  
We assume log-flat priors for semimajor axis $a$, primary mass $\mathrm{M_{\rm pri}}$, companion mass $\mathrm{M_{\rm sec}}$, and radial velocity jitter $\mathrm{\sigma_{jit}}$; a prior of sin$\textit{i}$ for inclination; and uniform priors for all other fitted parameters. Alternatively, the prior on the primary mass and its uncertainty can be set in $\tt [priors\_settings]$ to incorporate other knowledge of the star. 

In addition to radial velocity fitting, we typically fit orbits using both absolute and relative astrometry (though only absolute astrometry is strictly required). Relative astrometry can be utilized by providing the data path to the relative astrometry file containing the astrometric epochs from different high-resolution imaging instruments. For epoch astrometry, the Hundred Thousand Orbit Fitter \citep[\texttt{htof}][]{Brandt+Michalik+Brandt+etal_2021} 
package mentioned in Section \ref{subsec:absolute_relative_astrometry} is employed in \codename to provide parameter fits to the astrometric intermediate data from either \hipparcos data reduction or from \gaia. In order to use epoch astrometry, the file path to observational epochs and scan angles for \gaia, or to the intermediate data  of \hipparcos (original or re-reduction) must be provided. The \gaia { predicted scan epochs and angle} can be queried using the Gaia Observation Forecast Tool, publicly available at  $\mathrm{https://gaia.esac.esa.int/gost/index.jsp}$. \texttt{htof} parses the given intermediate data and extracts (inverse-)covariance matrices and epochs of observations to compute synthetic \hipparcos and \gaia catalog positions and proper motions described in Section \ref{subsec:astrometry}. 

The output of \codename is a single FITS file with {the chain and other calculated parameters in a FITS table in header-data unit (HDU) 1}.
The header of the first extension (HDU 0) contains the configuration parameters read from the \texttt{.ini} file and used to construct the chain.  
{The FITS table in HDU 1 contains columns with names, arrays, and units (where appropriate) for each fitted parameter, for the natural logarithm of the likelihood, and for other quantities computed at each step of the chain.  Each array has dimensions ${\tt nwalkers} \times {\tt nstep//thin}$.}

\begin{deluxetable*}{lccr}
\def\arraystretch{0.78}
\tabletypesize{\footnotesize}
\tablewidth{0pt}
\tablecaption{Description of the configuration file contents. \label{tab:configfile}}
\tablehead{
    \colhead{Parameter Name} &
    \colhead{Data Type} &
    \colhead{Default} &
    \colhead{Description}
    }
\startdata
{\tt [data\_paths]} \\ \hline
${\tt HipID}$               & Integer        & {\tt 0} & \hipparcos number.  If valid, load HGCA absolute astrometry \\
${\tt HGCAFile}$            & Data File     & [required] & The \hipparcos-\gaia Catalog (either the DR2 or EDR3 edition) \\
${\tt RVFile }$             & Data File     & {\tt ''} & File containing the radial velocity time series for the star \\
${\tt AstrometryFile}$      & Data File     & {\tt ''} & File containing the relative astrometry for the companion(s) \\
${\tt GaiaDataDir}$         & Directory     & {\tt ''} & Path to the \gaia scans as output by GOST in \texttt{.csv} format \\
${\tt Hip1DataDir}$         & Directory     & {\tt ''} & Path to the \hipparcos (original reduction) intermediate data \\
${\tt Hip2DataDir}$         & Directory     & {\tt ''} & Path to the \hipparcos (re-reduction) intermediate data\\
${\tt start\_file}$         & Data File     & {\tt 'none'} & File with the initial parameter guesses. If {\tt 'none'}, use default guesses \\[0.00em] \hline
{\tt [mcmc\_settings]} \\ \hline
${\tt ntemps}$              & Integer       & {\tt 10} & Number of temperatures to use in the parallel tempering chain \\
${\tt nwalkers}$            & Integer       & {\tt 100} & Number of walkers; each with {\tt ntemps} number of chains \\
${\tt nplanets}$            & Integer       & [required] & Number of companions to fit \\
${\tt nstep}$               & Integer       & [required] & Number of steps contained in each chain \\
${\tt thin}$ & Integer & {\tt 50} & Thinning of the chain (keep every ${\tt thin}$ step) \\
${\tt nthreads}$            & Integer       & {\tt 1} & Number of threads to use (the built-in parallelization is poor) \\
${\tt use\_epoch\_astrometry}$& Boolean       & {\tt False} &  Use the epoch astrometry in \texttt{GaiaDataDir}, \texttt{Hip1DataDir}, etc.? \\ 
{\tt jit\_per\_inst} & Boolean & {\tt False} & Fit a separate RV jitter for each instrument? \\[0.00em] \hline
{\tt [priors\_settings]} \\ \hline
${\tt mpri}$                & Float         & {\tt 1} & Mean of a Gaussian prior on primary mass (in $M_\odot$) \\
${\tt mpri\_sig}$            & Float         & {\tt inf} & Uncertainty in the stellar priors.  If {\tt inf}, use the default $1/M$ prior \\
{\tt minjitter} & Float & {\tt 1e-5} & Minimum allowable RV jitter (m/s).  Should be $>0$ \\
{\tt maxjitter} & Float & {\tt 1e3} & Maximum allowable RV jitter (m/s). Must be $> {\tt minjitter}$ \\
{\tt m\_secondary0} & Float & {\tt 1} & Mean (in $M_\odot$) of the (Gaussian) mass prior for companion 0 \\
{\tt m\_secondary0\_sig} & Float & {\tt 1} & Standard deviation of the (Gaussian) mass prior for companion 0 \\
{\tt parallax} & Float & {\tt None} & Parallax prior: required if star is not in the HGCA, ignored otherwise \\
{\tt parallax\_error} & Float & {\tt None} & Parallax prior if star is not in the HGCA \\[0.00em] \hline
{\tt [secondary\_gaia]}  & & & \gaia data of the secondary (set {\tt companion\_ID\,=\,-1} if undetected)\\ \hline
{\tt companion\_ID} & Integer & {\tt -1} & ID of the \gaia companion, should match entry in {\tt AstrometryFile} \\
{\tt pmra} & Float & {\tt 0} & \gaia proper motion (RA) of companion \\
{\tt pmdec} & Float & {\tt 0} & \gaia proper motion (Dec) of companion \\
{\tt epmra} & Float & {\tt 1} & \gaia proper motion uncertainty (RA) of companion \\
{\tt epmdec} & Float & {\tt 1} & \gaia proper motion uncertainty (Dec) of companion \\
{\tt corr\_pmra\_pmdec} & Float & {\tt 0} & Correlation ($\in (-1, 1)$) between {\tt pmra} and {\tt pmdec} \\[0.00em] \hline
{\tt [plotting] } \\ \hline
${\tt McmcDataFile}$        & Data File     & [required] & Path to MCMC chain produced from the $\tt fit\_orbit$ command \\
${\tt burnin}$              & Integer         & {\tt 0} & Burnin length for thinned chains \\
${\tt check\_convergence}$              & Boolean         & {\tt False} & Make diagnostic plots to help check for convergence? \\
{\tt iplanet}  & Integer & {\tt 0} & ID of the companion to plot ($0 \leq {\tt iplanet} < {\tt nplanets}$) \\
${\tt target}$              & String        & {\tt ''} & Name of the target, used for file naming \\
${\tt start\_epoch}$        & Integer       & {\tt 1950} & Customized range of dates (fractional years) \\
${\tt end\_epoch}$          & Integer       & {\tt 2030} & Customized range of dates (fractional years) \\ 
${\tt num\_orbits}$         & Integer       & {\tt 50} & Number of random orbits drawn from the posterior distribution \\
${\tt num\_steps}$          & Integer       & {\tt 1000} & Points per plotted orbit (aliasing can occur if {\tt num\_steps} is too small) \\
${\tt predicted\_years}$    & Float(s)      & {\tt 2010,2020} & Labeled year(s) on the best-fit orbit in the {\tt Astrometry} plot \\
${\tt position\_predict}$    & Float       & {\tt 2020} & Epoch (fractional year) for the {\tt Astrometric\_prediction\_plot} \\ 
${\tt Astrometry\_orbits\_plot}$         & Boolean         & {\tt True}  & Plot the astrometic orbits? \\
${\tt Astrometric\_prediction\_plot}$            & Boolean           & {\tt True} & Plot the density plot for the predicted epoch? \\
${\tt RV\_orbits\_plot}$                    & Boolean           & {\tt True} & Plot the full RV orbits? \\
${\tt RV\_plot}$                    & Boolean           & {\tt True} & Plot RV orbits vs.~epoch and O-C over the baseline of RV data? \\
${\tt RV\_Instrument}$                    & Integer/String           & {\tt All} & ${\tt All}$ or Instrument number \\
${\tt Relative\_separation\_plot }$         & Boolean           & {\tt True} & Plot relative separation vs. epoch and O-C? \\
${\tt Position\_angle\_plot}$               & Boolean           & {\tt True} & Plot position angle vs. epoch and O-C? \\
${\tt Proper\_motion\_plot}$                & Boolean           & {\tt True} & Plot the proper motions in RA and Dec and O-C? \\
${\tt Proper\_motion\_separate\_plots}$                & Boolean           & {\tt False} & True if two separate plots for the proper motions are desired \\
${\tt Corner\_plot }$                       & Boolean           & {\tt True} & Plot a two dimensional corner plot from MCMC chain? \\ 
${\tt set\_limit}$                          & Boolean           & {\tt False} & Use user-specified axis limits? \\
${\tt xlim }$                            & Floats             & {\tt None} & If ${\tt set\_limit}$ is True, set x limits with two comma-separated values \\
${\tt ylim}$                                & Floats             & {\tt None} & If ${\tt set\_limit}$ is True, set y limits with two comma-separated values\\
${\tt marker\_color}$                       & String & {\tt blue} & Matplotlib color of the marker for the observed data points \\
${\tt use\_colorbar}$                       & Boolean           & {\tt True} & Turn on/off colorbars. \\
${\tt colormap}$                            & String  & {\tt viridis} & Colormap name from the Matplotlib colormap library \\
${\tt reference}$                           & String            & {\tt msec\_jup} & Colormap reference, ${\tt msec\_jup}$, ${\tt msec\_solar}$ or ${\tt ecc}$.  \\
${\tt show\_title}$                         & Boolean           & {\tt False} & Turn on/off the title of the plot \\
${\tt add\_text}$                           & Boolean            & {\tt False} & True if adding {\tt text\_name} somewhere on the plot \\
${\tt text\_name}$                          & String            & {\tt None} & If ${\tt show\_title}$ is True, specify the text \\
${\tt x\_text}$                             & Float             & {\tt None} & If ${\tt add\_text}$ is True, enter the x coordinate of the text \\
${\tt y\_text}$                             & Float             & {\tt None} &If ${\tt add\_text}$ is True, enter the y coordinate of the text \\[0.00em] \hline
{\tt [save\_results] } \\ \hline
${\tt save\_params}$        & Boolean     & {\tt True} & Save the posterior parameters to a .txt file? \\
${\tt err\_margin}$        & Float(s)      & {\tt 0.16,0.5,0.84} &  Quantiles for posterior parameters and uncertainties 
\enddata
\end{deluxetable*}

We have implemented a plotting routine in Python to visualize the results obtained from the FITS file containing the MCMC chains produced by a joint orbit fit. In the $\tt [plotting]$ section of the {\tt .ini} configuration file, the file path to the FITS file and a $\tt burnin$ length for the (thinned) MCMC chains must be specified. Also, users have the option to select which plots to generate using $\tt True$ and $\tt False$ values for the plot keywords. Here, we briefly describe the plots that \codename is currently configured to produce. Section \ref{sec:casestudy} will provide an example fit and sample plots.

\subsection{Astrometric orbits}

The relative astrometric orbits for companions of stellar or planetary systems can be generated by setting $\tt Astrometry\_orbits\_plot$ to $\tt True$. Several features are worth mentioning in this plot. The thick black line indicates the highest likelihood orbits; thin lines are orbits randomly drawn from the posterior distributions and colored according to either the companion mass or eccentricity. This is the case for all the plots described below, except for the astrometric prediction plot which is a 2D contour.  The black dashed line inside the most-likely orbit is the line of nodes joining the ascending node and the descending node of the most-likely orbit. This line of nodes indicates the position of the orbital plane of the system with respect to the sky plane. The unfilled circles plotted along the most-likely orbit are the predicted positions of the companion at specific user-defined epochs from $\tt predicted\_years$. If an astrometric file is provided from $\tt AstrometryFile$, the observations will be plotted as filled circles.  If there is more than one companion included in the fit, the user may specify which one is plotted using the keyword {\tt iplanet} (with $0 \leq {\tt iplanet} < {\tt nplanets}$).

\subsection{Astrometic prediction}

\codename can predict the location of a companion relative to its host star at a specified epoch.  The resulting density plot 
can be obtained by setting $\tt position\_predict$ to the desired epoch, and setting $\tt Astrometric\_prediction\_plot$ to $\tt True$. This contour plot shows the posterior probability density of the predicted positions of the companion in terms of relative offsets from the primary star in right ascension and declination, with the inner contour being the most likely position of the companion at that future epoch. The 1-$\sigma$, 2-$\sigma$, and 3-$\sigma$ contours enclose 68.3\%, 95.4\%, and 99.7\% of the posterior probabilities for the future location. 
    
\subsection{Radial velocity orbits}

\codename can make two plots of the stellar radial velocity: one restricted to the observational time frame, and one spanning a longer, user-specified range of dates to show longer-term behavior.  To make the latter plot
the user may set 
$\tt RV\_orbits\_plot$ to $\tt True$. {The former plot, restricted to the observational baseline, 
requires the user to give the path for $\tt RVFile$. The $\tt RVFile$ containing the observed radial velocity time series must have the format specified by \autoref{tab:fileformats}.} 
The filled circles on {both plots} represent RV data from all the radial velocity instruments. The observed RV data are shifted by an offset according to the maximum likelihood zero point for each RV instrument; {the maximum likelihood offsets for each instrument are given by fields named, e.g., {\tt `RV\_ZP\_0\_ML'} for instrument 0.} 
    
\codename can also make a plot of the RVs
restricted to the time range sampled by the RV instruments; $\tt RV\_plot=True$ enables this plot. 
This is essentially a zoomed-in view of the part of the RV orbits plot described above.  It contains RV data from single- or multi-instrument RV observations, plus a corresponding Observed$-$Calculated (hereafter O$-$C) residual shown 
underneath. To choose to plot the RVs from a specific or from all the RV instruments, users may set $\tt RV\_Instrument$ to the instrument number or to {\tt `All'}. The observed data are represented by the solid circles with error bars. The O$-$C residual indicates both the deviation of the observed value from the most-likely orbit and the variation in RV across the orbits randomly drawn from the posterior. The error bars include the RV jitter of the best-fit orbit. 
    
\subsection{Relative separation and position angle}
For relative astrometry, \codename offers two more plots in addition to the astrometric orbits plot: the relative separation (in arcseconds) and the position angle (in degrees) of the imaged companion relative to the primary star in a time range sampled by the relative astrometric instruments, and their corresponding O$-$C residuals. These two plots can be generated if the user sets $\tt Relative\_separation\_plot$ or $\tt Position\_angle\_plot$ to {\tt True}. In each case, the  path to $\tt AstrometryFile$ should be specified. The orbit's relative separation in arcseconds is the product of the projected relative separation $\rho$ in AU and the parallax $\varpi$; we use the best-fit parallax for each set of orbital parameters. We refer the reader to Section \ref{subsec:absolute_relative_astrometry} for a detailed description of the equations, and to the Appendix for the calculation of the best-fit parallax. The solid circles with error bars indicate the relative astrometric data from direct imaging instruments. Due to each instrument's having different uncertainties, data reduction methods, and variations in the field rotation and plate scale, users may choose to adjust the imaging data in $\tt AstrometryFile$ to keep or discard a set of imaging data.  Only one companion will be plotted.  If there is more than one companion fit, the user may specify which one to plot using {\tt iplanet}.   
    
\subsection{Proper motion}

Variations in proper motion induced by the companion(s) on the primary star as measured from absolute astrometry from \hipparcos and \gaia can be plotted with $\tt Proper\_motion\_plot=True$. \codename computes these proper motions as instantaneous values using the time derivative of the eccentric anomaly. In contrast, both \hipparcos and \gaia fit sky paths to the star's {\it position} as measured over several years.  As a result, the observed data points and error bars do not correspond to the plotted lines in the same way that they do for, e.g., radial velocity.  \codename does not plot the \hipparcos-\gaia mean proper motion, as this measurement does not represent an instantaneous proper motion at the mean epoch.  It is, rather, a measurement of the integral of the proper motion.  The O$-$C curves should be taken with caution.  The actual $\chi^2$ values for each set of measurements (which are meaningful and also include measurement covariance) are available in the 
FITS file written by \codename.

To plot the two components of the proper motion as two separate, individual plots (one for right ascension and one for declination), users may set $\tt Proper\_motion\_separate\_plots$ to {\tt True}. The data with error bars are from the cross-calibrated absolute astrometry of {the} HGCA. {The} HGCA provides three proper motions: one near 1991.25, another near 2015.5, and the positional difference between the \hipparcos re-reduction and the \gaia catalog {(either the DR2 or EDR3 edition)} scaled by the time between them. \codename plots data points for only the \hipparcos and \gaia proper motions.  \codename adds the best-fit barycenter proper motion, calculated as described in the appendix, to each orbit.  

\subsection{Corner plot}

A two dimensional corner plot of the MCMC chain can be obtained by setting $\tt Corner\_plot$ to True. A $\tt burnin$ phase for the thinned MCMC chain can be specified for plotting. We have modified $\tt corner.py$ from \cite{2016JOSS....1...24F} to format the titles displayed on top of the histograms in this plot. We chose to keep two significant figures in the uncertainties with a matching number of decimal places in the values.

The corner plot shows astrophysically meaningful Keplerian orbital elements including semi-major axis $a$, eccentricity $\varepsilon$, and inclination $i$, and the masses of the primary star in $\Msun$ and its companion in $\Mjup$ from the joint RV and astrometric MCMC analysis. If more than one companion is fit, the corner plot will use the parameters from the companion designated by {\tt iplanet}.

\section{Case study: application to HD 159062B}
\label{sec:casestudy}

\begin{deluxetable*}{llllllll}
\tablecaption{HGCA proper motions\label{tab:HGCA_pm}
}
\tablewidth{0pt}
      \tablehead{
        \colhead{\bf Source} &
        \colhead{$\mu_{\alpha*}$~(mas\,yr$^{-1}$)}  &
        \colhead{$\mu_{\delta}$~(mas\,yr$^{-1}$)} &  
        \colhead{\bf corr}  &   
        \colhead{$t_{\alpha*}$}   & 
        \colhead{$t_{\delta}$} & 
        \colhead{$\varpi$ (mas)} &
        \colhead{RV (km\,s$^{-1}$)}
        }
        \startdata
\hipparcos &     $174.316 \pm 0.666$ &   $75.598 \pm 0.612$  &0.27&  1991.20&  1991.12 & \\
HG    &  $172.499 \pm 0.019$    & $75.776 \pm 0.020$ & 0.11 & \\
\gaia EDR3 &   $169.814  \pm 0.026$  &  $77.133 \pm 0.029$ & 0.22 & 2016.07 & 2016.27 & $46.118 \pm 0.024$ & $-84.11 \pm 0.18$ 
\enddata
\end{deluxetable*}
\vspace*{+10mm}

In this section, we provide a case study application of \codename to a nearby white dwarf/main sequence (WD/MS) binary system, HD 159062. The white dwarf was discovered by \cite{Hirsch_2019}, hereafter H19, who fit for its orbit and mass.  We first review the system, a widely-separated binary with a degenerate companion to a main sequence star, before summarizing the results of H19.  We then discuss our own orbital fit and its implications for the binary system's past evolution.

\subsection{Background on HD~159062}

Main sequence stars with masses $\lesssim$8 $\Msun$ will evolve off the main sequence through the asymptotic giant branch (AGB) phase at the stellar evolutionary end-point and cool over billions of years until they eventually become dense white dwarfs \citep{Dobbie_2006}. More than 97\% of stars are expected to evolve into white dwarfs \citep{Renedo_2010}. 
White dwarfs are used as powerful tools to trace the evolution of the Galaxy, and to provide constraints on global stellar populations. White dwarfs in binaries can also be potential progenitors of Type 1a supernovae \citep{Hillebrandt+Niemeyer_2000}. 

Spectroscopic and photometric surveys such as SDSS, Pan-STARRS and \gaia can reveal fundamental properties of white dwarfs including mass, cooling rate and age, and atmospheric and internal composition. The recent \gaia DR2 catalog has provided updated precise astrometric and photometric data that can be used to infer the local white dwarf population \citep{Fusillo_2019}. Since most stars end their lives as WDs and most stars reside in binaries \citep{Moe_2017}, it is estimated that around one quarter of the more than two hundred known white dwarfs within 25~pc of the Sun reside in WD/MS or WD/WD binary systems \citep{Holberg_2016}. Population synthesis modeling shows this number is reasonable but the observed rate of WD/MS binaries likely suffers from selection effects due to low detection sensitivity to faint white dwarfs near bright main sequence stars \citep{Toonen_2017}. 

The host star of the system we study here, HD~159062A, is revealed spectroscopically as an old main sequence G-K dwarf star with low metallicity based on its $\log g$ and $T_{\rm eff}$ (H19). The observed Ca II HK emission $R^{'}_{\rm HK}$ and rotation period 
led to an age diagnostic of $\sim$ 7 Gyr. Literature reports on HD 159062A's age as derived from high-resolution spectroscopy range from approximately 5 Gyrs (e.g.~\citet{Torres+Cai+Brown_2019}; \citet{Raghavan_2012}) to 8 Gyrs (e.g.~\citet{Pace_2013}) to 14 Gyrs (e.g.~\citet{luck_2017}; \citet{bai_2018}). 

\textcolor{black}{Before the discovery of the white dwarf HD 159062B, \citet{Fuhrmann_2017a} predicted the existence of a degenerate companion around HD 159062A due to a barium overabundance of $[{\rm Ba/Fe}] = +0.4$ dex in HD 159062A. This is based on an empirical relation of the barium-to-iron enrichment as a function of age derived from an all-sky local population study of ancient Population II and intermediate-disk stars. 
Deviants from this Ba/Fe ratio versus age relation included five known stars with degenerate companions; HD 159062A was the most extreme outlier with an overabundance of barium. These five known systems almost certainly underwent mass transfer. \citet{Fuhrmann_2017a} characterize them as field blue stragglers
due to high rotation, excessive chromospheric activity, lithium depletion, low orbital eccentricity and noticeable age discrepancies with activity-based age indicators \citep{Fuhrmann_1999}. }

\textcolor{black}{
Solar-type Population II stars are old; they rotate at low rotational velocities and show low levels of chromospheric and coronal activity \citep{Barnes_2003,Mamajek+Hillenbrand_2008}.  Additional observed chromospheric activity could be explained by mass and angular momentum transfer from binary companions. Many of the stars with degenerate companions studied by \citet{Fuhrmann_2017a} had slight barium depletions and $\lesssim$1000-day orbital periods, suggestive of mass transfer from first-ascent red giant progenitors.  
HD 159062, however, had an 
anomalously high Ba/Fe ratio unlikely to have 
arisen from a field anomaly at its formation epoch; it must have been instead due to the s-process nucleosynthesis product of a binary companion.}

\textcolor{black}{
HD 159062's overabundance of barium suggests wind accretion of s-process elements from a more distant AGB progenitor rather than Roche lobe overflow from first-ascent red giant progenitor \citep{Boffin+Jorissen_1988,han_1995,Jeffries_1996}, so HD 159062B must have a longer period than $\sim$1000 days. Indeed, \citet{Fuhrmann_2017a} argue that barium enrichment via wind accretion only works for orbital periods from about 10 to 1000 years. This claim is backed up by an example of HD 114174, a cool white dwarf companion separated by $59.8 \pm 0.4 $AU from the G-type host star, whose orbital period is estimated to be between 154 and 881 years for eccentricities in the range 0$\leq$ e $\leq$ 0.5 \citep{Matthews_2014}. A barium overabundance of $[{\rm Ba/Fe}] = +0.24$ dex is observed of the G star. The projected separation excludes the possibility of any substantial past mass exchange via Roche lobe overflow, however, \citet{Matthews_2014} speculated that its orbit may have been pushed significantly outward during dynamical events such as the ejection of a third body.}

\textcolor{black}{The case of HD 114174 suggests a similar scenario for HD 159062, and the presence of a wide, white dwarf companion.  HD 159062B could be located at a wider separation of tens of AUs away from HD 159062A whose observed barium overabundance can be explained by wind accretion of s-process materials from a former AGB primary that survived as a white dwarf companion, as well as any other dynamical events that may have widened its orbit. Apart from HD 114174 and HD 159062, \citet{Fuhrmann_2017a} also identified two systems, HR 3578 and 104 Tau, from their Ba/Fe ratio studies, both of which are without known companions but with slight barium anomalies. However, both stars' lack of strong astrometric accelerations are not suggestive of any orbiting white dwarf companions at orbital periods of 10 - 1000 years. HR 3578 does show a $\sim$3$\sigma$ acceleration in declination in the HGCA.  104~Tau has been reported to be a visual binary \citep{Eggen_1956}, but more recent interferometric non-detections \citep{Raghavan+McAlister+Henry+etal_2010}, and stable radial velocities with a shallow trend \citep{Butler+Vogt+Laughlin+etal_2017} strongly suggest otherwise.  HR 3578 and 104~Tau both represent good targets for follow-up imaging.}

\subsection{Data and Previous Fit}

HD 159062B was discovered by H19 in 14 years of precise radial velocity (RV) data from the HIRES spectrograph on Keck \citep{Vogt+Allen+Bigelow+etal_1994,Howard+Johnson+Marcy+etal_2010}, and imaged using multi-epoch multi-band imaging observations from the ShaneAO system on the Lick Observatory 3-meter Shane telescope \citep{Gavel+Kupke+Dillon+etal_2014,Srinath+McGurk+Rockosi+etal_2014}, the PHARO AO system on the Palomar Observatory 5-meter telescope \citep{Hayward+Brandl+Pirger+etal_2001}, and the NIRC2 AO system at the Keck II 10-meter telescope \citep{Wizinowich+Acton+Shelton+etal_2000}. H19 performed a joint MCMC analysis using 45 radial velocities and three NIRC2 astrometric measurements to derive the best orbital parameters. They assumed Gaussian priors on the mass of the primary star, the difference between the pre-upgrade velocity zero point and the post-upgrade velocity zero point and parallax, and uniform priors on all other fitted parameters. H19 derived a new spectroscopic mass of $0.76 \pm 0.03 \Msun$, but 
adopted a prior of $0.80 \pm 0.05 \Msun$ based on literature values. H19 used 10 temperatures and 300 walkers in their MCMC in a total of $10^5$ steps.
After the initial MCMC analysis, an additional prior on the white dwarf cooling age was added based on white dwarf cooling models and photometric constraints from ShaneAO, PHARO, and NIRC2s' photometry measurements in $J$, $K_{s}$ and $L'$. H19 concluded that 
the companion is an old $\mathrm{M_{B} = 0.65^{+0.12}_{-0.04} \Msun}$ with an orbital period of $P = 250^{+130}_{-76}$ years, and a cooling age of $\mathrm{\tau = 8.2^{+0.3}_{-0.5}}$ Gyr. 

{More recently, \cite{Bowler+Cochran+Endl+etal_2021} performed an orbital fit using similar data and \codename with the DR2 version of the HGCA.  Our results are similar to theirs, but are slightly more precise with the updated absolute astrometry from \gaia EDR3.  We omit the additional RVs from \cite{Bowler+Cochran+Endl+etal_2021} (which do not extend the observational baseline) and the additional relative astrometry.  This enables a direct comparison of our results with those of H19 and shows the power of absolute astrometry to constrain the orbit of HD 159062AB.}

\subsection{A New Fit}

We use \codename to perform a comprehensive joint MCMC analysis of HD 159062B using the same RV and imaging data as H19, but adding the cross-calibrated absolute astrometry of the HGCA. 
\autoref{tab:HGCA_pm} summarizes the HGCA proper motions for HD 159062B, including the correlation coefficients between proper motions in right ascension and declination, and the central epoch for each measurement. The relative astrometry data are from three direct imaging instruments: ShaneAO, PHARO and NIRC2. Similarly to H19, we restrict our analysis to NIRC2 with its well-measured distortion correction and track record of precision astrometry \citep{Service_2016,Sakai+Lu+Ghez+etal_2019}.

\begin{deluxetable}{ll}
\tablewidth{0pt}
\tablecaption{Basic Parameters and Default Priors\label{tab:priors}}
\tablehead{
        {\bf Parameter} & 
        {\bf Prior}
        }
\startdata
RV Jitter  $\sigma_{\rm jit}$     &  $1/\sigma$ (log-flat) \\
Primary Mass $M_{\rm pri}$        &  $1/M$ (log-flat) \\
Secondary Mass $M_{\rm sec}$      &  $1/M$ (log-flat) \\
Semimajor axis $a$                &  $1/a$ (log-flat) \\
$\sqrt{\varepsilon} \sin \omega$  &  uniform \\
$\sqrt{\varepsilon} \cos \omega$  &  uniform \\
Inclination $i$                   & $\sin(i)$, 0$^{\circ}<i<180 ^{\circ}$ \\
Mean longitude at 2010.0 $\lambda_{\rm ref}$ &                         uniform \\
Ascending node $\Omega$           &  uniform \\
Parallax $\varpi$                 &  $\exp[-\frac{1}{2} (\varpi - \varpi_{\it Gaia})^2/\sigma_{\varpi,\it Gaia}^2 ]$ 
\enddata
\end{deluxetable}

Using \codename, we ran 10 temperatures and 100 walkers over $10^{5}$ steps to fit for nine parameters, keeping every 50$^{\rm th}$ step.
\textcolor{black}{Our results are based on the `coldest' of 10 chains, with the `hottest' chain being the chain that effectively samples all of the allowed parameter space. The total time taken using two cores was 2177 s $\sim$ 0.59 hr, and the mean acceptance fraction (cold chain) was 
0.074.}
\autoref{tab:priors} provides a complete list of the parameters that we fit and the priors we used. We have marginalized out four parameters: the parallax $\varpi$, proper motion of the system barycenter, and RV offset as described in Section \ref{sec:likelihood} and the Appendix. 

\begin{deluxetable*}{lll}
\tablewidth{0pt}
\tablecaption{MCMC Results}
\tablehead{
        {\bf Parameter} & 
        {\bf Median$\pm 1\sigma$} & 
        {\bf 95.4\% C.I. }  
}
\startdata
\multicolumn{3}{c}{Fitted parameters} \\ \hline
RV Jitter $\sigma_{\rm jit}$ $({\rm m\,s}^{-1})$  &   ${1.26}_{-0.30}^{+0.32}$        &  (0.65, 1.917) \\
Primary Mass $M_{\rm pri}$ $(M_{\odot})$          &   $0.80 \pm 0.05$     &  (0.70, 0.90) \\
Secondary Mass $M_{\rm comp}$ $(M_{\odot})$       &   ${0.608}_{-0.0073}^{+0.0083}$   &  (0.594, 0.625) \\
Semimajor axis $a$ (AU)                           &   ${61.9}_{-7.2}^{+7.0}$          &  (46.9, 83.3)  \\
$\mathrm{\sqrt{e}\, sin\, \omega}$                &   ${-0.17}_{-0.11}^{+0.15}$       &  ($-$0.35, 0.13)\\
$\mathrm{\sqrt{e}\, cos\, \omega}$                &   ${0.00 \pm 0.32}$      &  ($-$0.52, 0.55) \\
$\mathrm{Inclination}~i~(^\circ)$                 &   ${63.0}_{-2.4}^{+1.8}$          &  (56.1, 66.6)  \\
Mean longitude at $\mathrm{\lambda_{ref}~(^{\circ})}$\tablenotemark{a}                                                                                                            &   ${151.5}_{-6.5}^{+8.0}$         &   (131.845, 167.314) \\
Ascending node $\mathrm{\Omega~(^{\circ})}$       &   ${133.4}_{-1.3}^{+1.7}$         &  (130.97, 138.92) \\ 
Parallax $\varpi$ (mas)                           &   $46.1856 \pm 0.0042$ &  (46.177, 46.194) \\
\hline
\multicolumn{3}{c}{Derived parameters}\\
\hline
Period (years)                                    &   ${411}_{-70}^{+71}$             &  (270, 641)	 \\
Argument of periastron $\omega\, (^{\circ})$      &   ${260}_{-76}^{+70}$             &  (40, 348)	 \\
Eccentricity $\varepsilon$                        &   ${0.102}_{-0.066}^{+0.11}$      &  (0.006, 0.37) \\
Semimajor axis $a$ (mas)                          &   ${2862}_{-330}^{+320}$          &  (2170, 3850) \\
Time of periastron $T_{0} = \mathrm{t_{ref}} - P\frac{\lambda - \omega}{360^{\circ}}$ (JD)                                                                                         &${2506737}_{-31268}^{+16428}$   & (2463393.772, 2583267.863)	 \\
Mass ratio                                        &   ${0.761}_{-0.045}^{+0.050}$     &  (0.677, 0.866) 
\enddata
\tablenotetext{a}{The reference epoch is $\mathrm{t_{ref} = 2455197.5\, JD\, (\lambda_{ref}; 2010\, Jan\, 1\, 00:00\, UT)}$. }
\label{tab:posteriors}
\end{deluxetable*}

We examine the chains of each parameter to make sure that the burn-in phase was complete and all walkers had stabilized in the mean and standard deviation of the posterior.\textcolor{black}{We discard the first 500 recorded steps (the first 25000 overall, as we save every 50$^{\rm th}$) as the burn-in phase so that the chains are ready to be used for inference.} The derived posterior probabilities from our joint orbit fit are given in \autoref{tab:posteriors}. With this case study, we provide an example of the eight plots produced by \codename, described in Section \ref{sec:configuration}. Two types of reference schemes are available for coloring the curves randomly drawn from the posterior distributions, either based on the mass of the secondary companion $M_{\rm sec}$ or eccentricity $\varepsilon$. We demonstrate both reference schemes for HD 159062B. 

\begin{figure*}
\includegraphics[height=.26\textwidth]{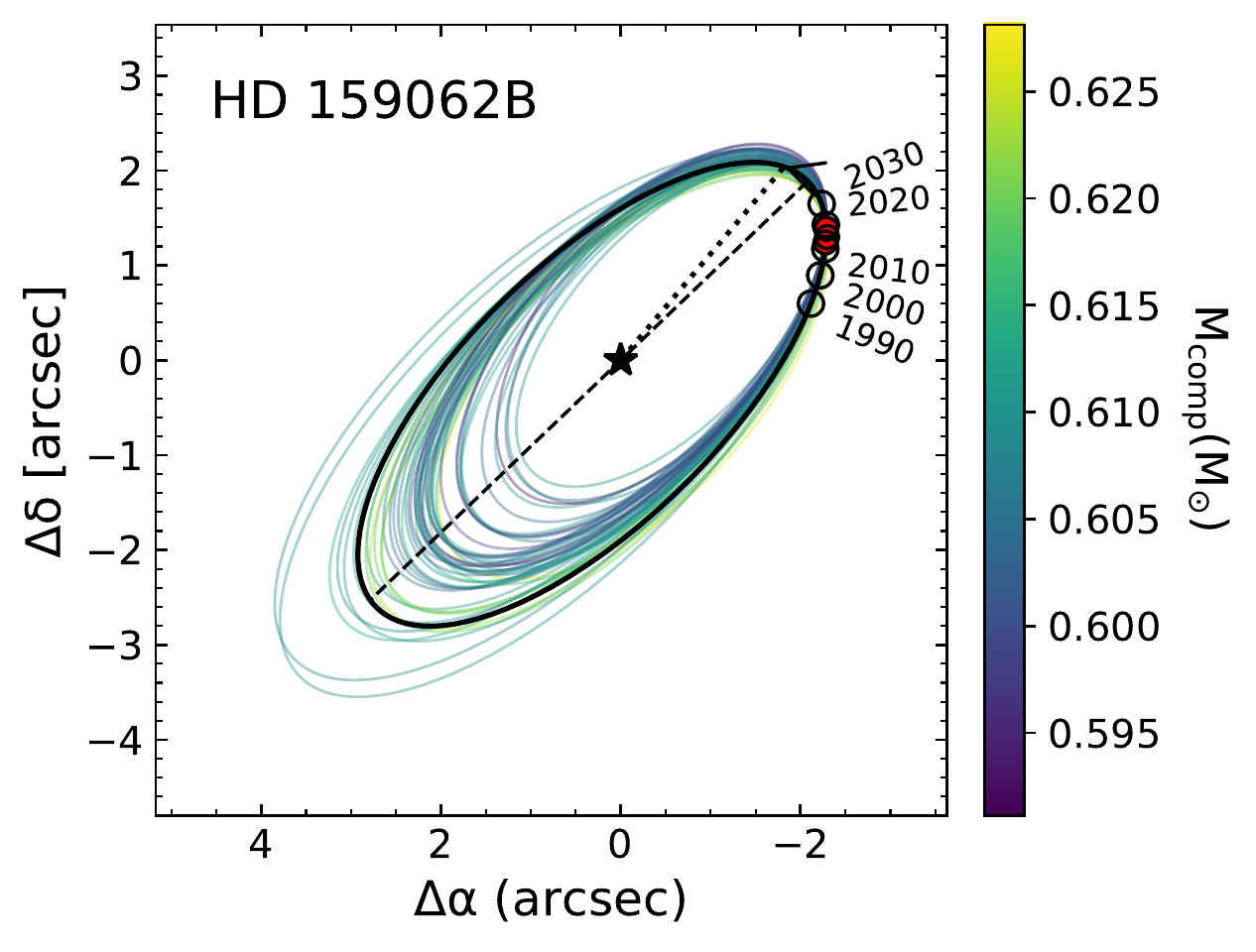}
\includegraphics[height=.26\textwidth]{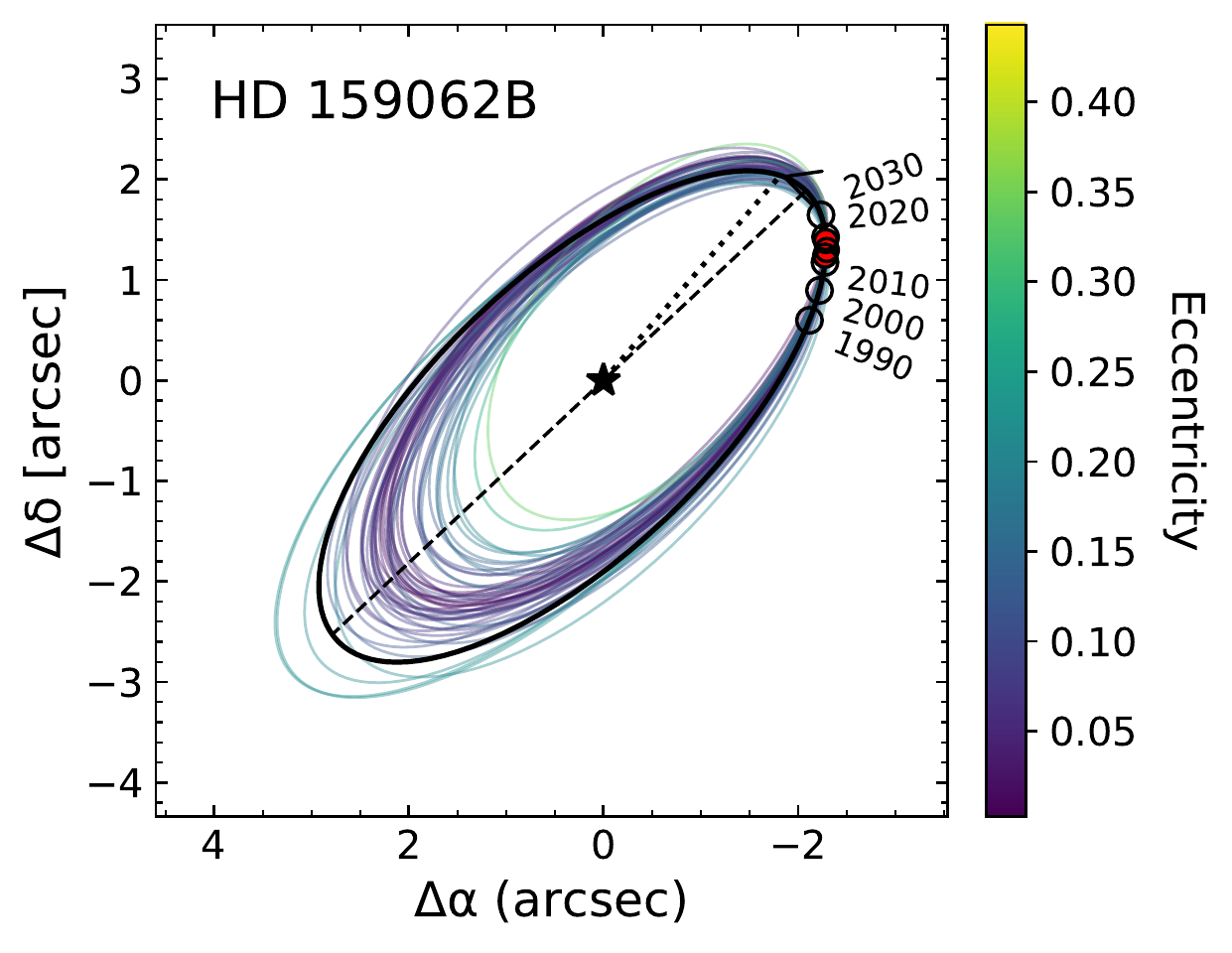} 
\includegraphics[height=.27\textwidth]{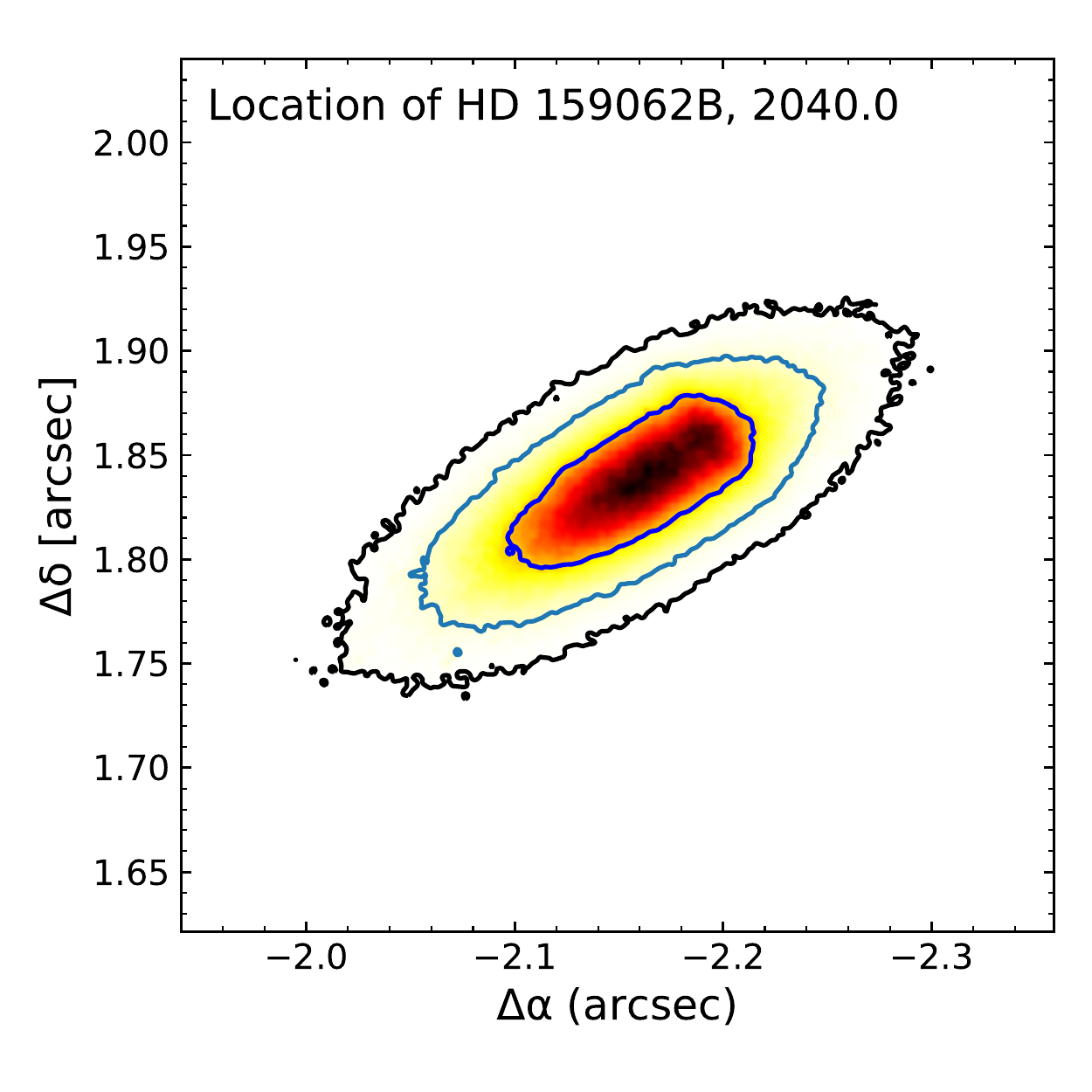}
\caption{Relative astrometric orbits and predicted future position for HD 159062B. Left: relative astrometric orbits colored according to the secondary mass. Middle: relative astrometric orbits colored by eccentricity. Right: astrometric prediction of the location of HD 159062B in 2040. $\bf Left\, and\, Middle:$ the thick black lines indicate the highest likelihood orbit and the colorful thin lines from purple to yellow are 50 orbits drawn randomly from the posterior distributions colored according to the companion mass (left panel) or eccentricity (middle panel). The dotted lines connect the host star to the periastron passages. The black empty circles along the maximum likelihood deprojected orbit indicate epochs spaced by 10 years from 1990 to 2030. The dashed lines are the line of nodes. The filled orange circles are plotted along the maximum likelihood at the epochs corresponding to the relative astrometry used in our analysis. The arrows mark the direction of motion of the secondary companion. $\bf Right:$ The contour lines demonstrate the likelihood of the location of HD 159062B in 2040, ranging from light yellow (least likely) to red to black (most likely).} 
\label{pics:astrometry_HD159062B}
\end{figure*}

\begin{figure*}
\centering
\includegraphics[width=.45\textwidth]{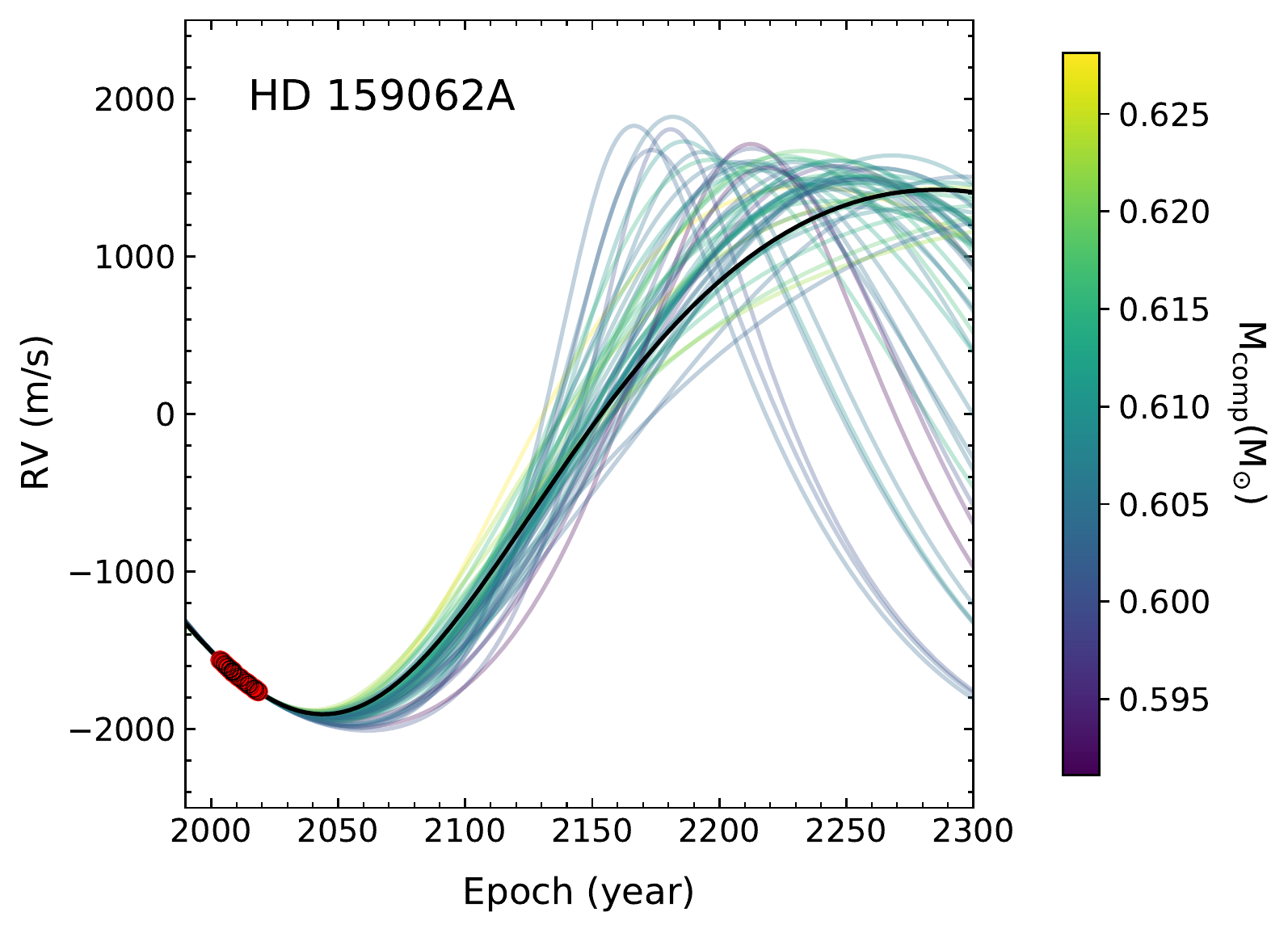}\quad
\includegraphics[width=.45\textwidth]{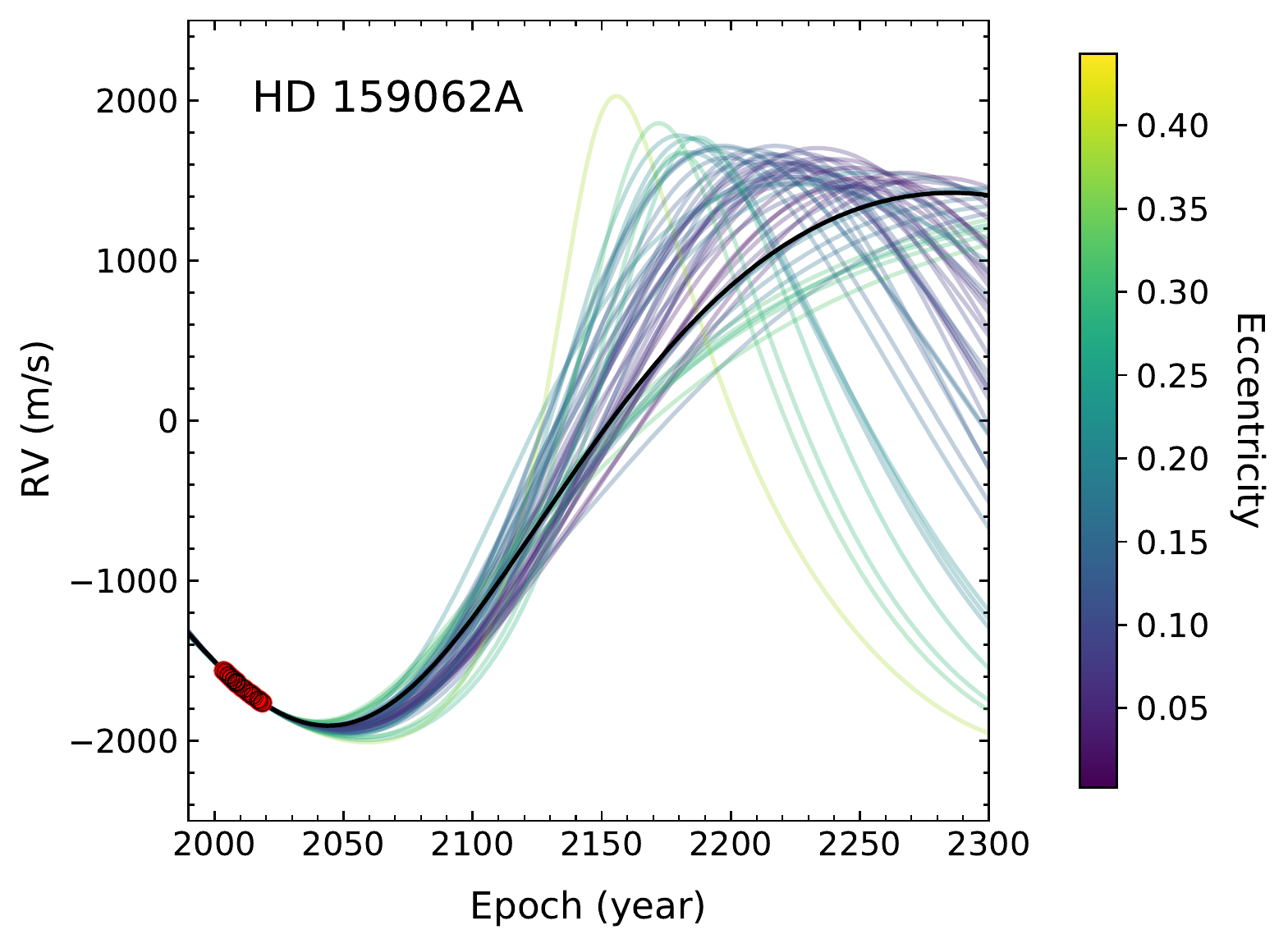}\quad 
\includegraphics[width=.45\textwidth]{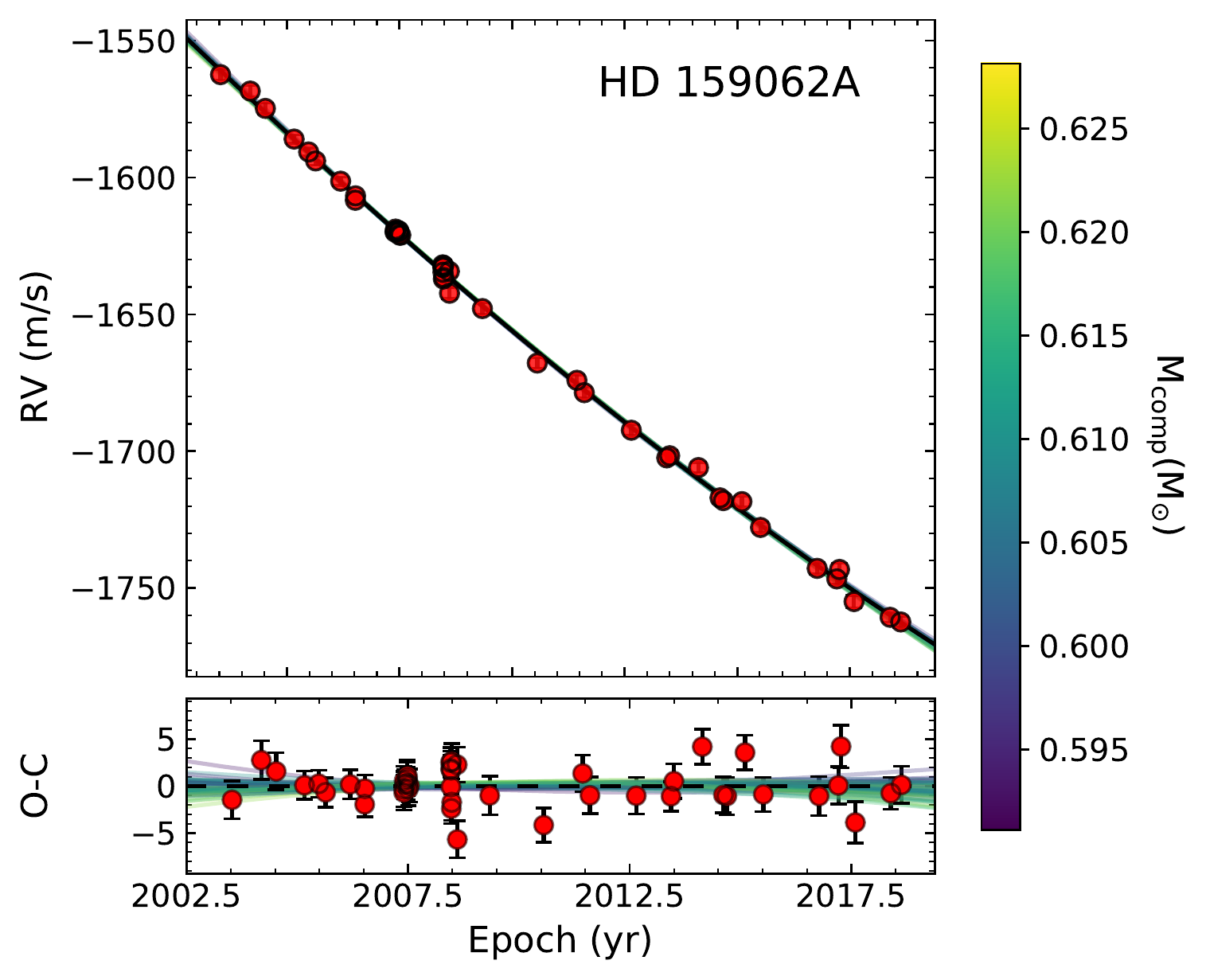}\quad
\includegraphics[width=.45\textwidth]{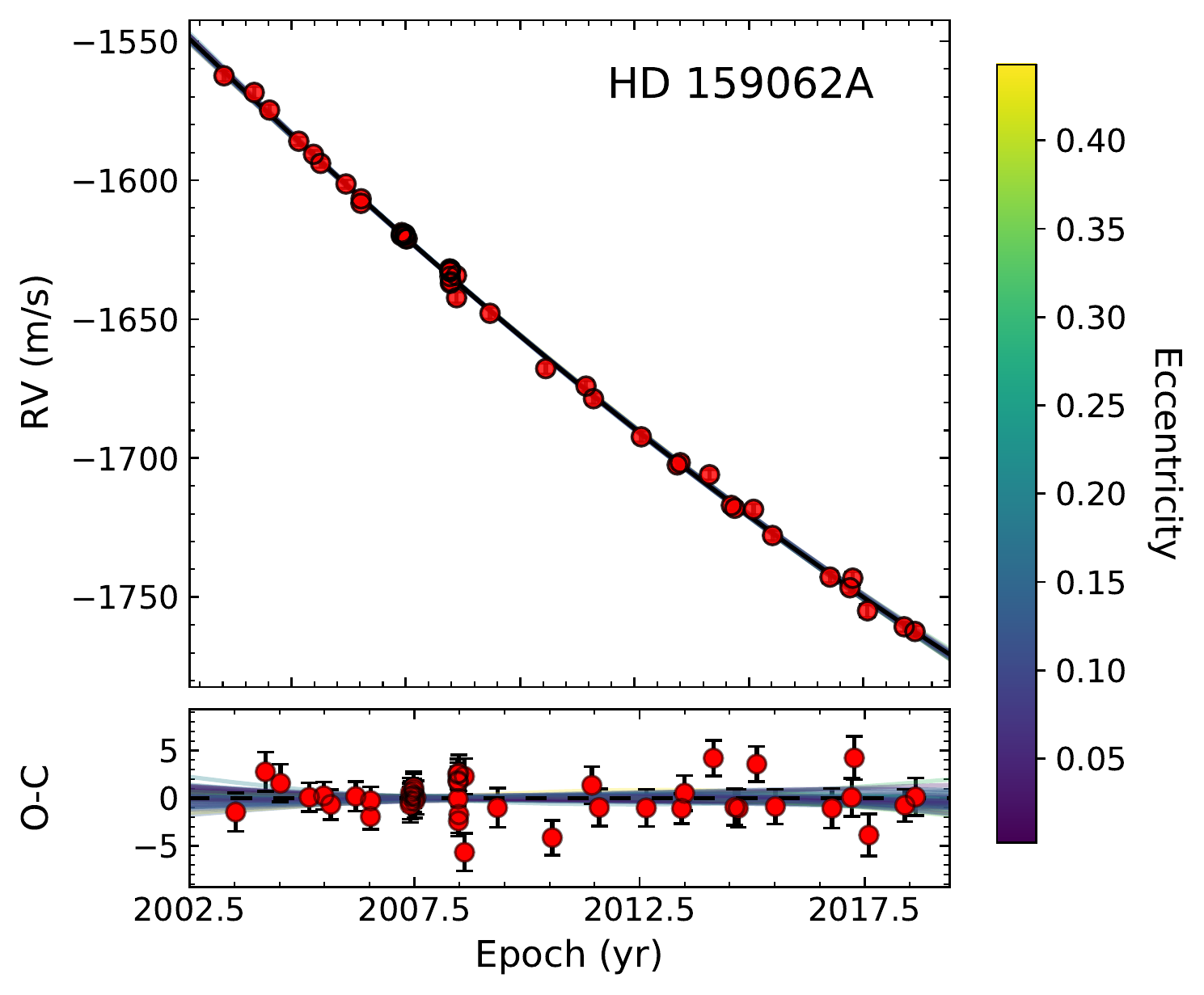}
\caption{{\bf Top:} RV orbits induced by the directly imaged companion over a significant fraction of its very long period, ranging from 1990 to 2280. {\bf Bottom:} RVs and the observed-calculated residuals, restricted to the observational time frame. In all of the panels, the thick black line is the highest likelihood orbit and the colorful lines are 50 orbits randomly drawn from the posterior probability distribution for HD 159062B. They are colored according to secondary mass (left panels) or eccentricity (right panels). The red solid points with error bars are RV data from Keck/HIRES with the best-fit RV zero point added. Error bars are too small to be visible except for some points on the plot of residuals. 
An RV of zero represents the system's barycentric velocity for the maximum likelihood orbit. 
The small bottom panel in each plot shows residuals after subtracting the RV orbit from the measurements. Due to the long orbital period of HD 159062B, more measurements are required to better constrain the orbital parameters. Moreover, we observe that with eccentricity as the reference coloring instead of the secondary mass, the RVs and residuals show more distinguishable structures.  Lower eccentricity (more purple) orbits agree better with the highest likelihood orbit.
}
\label{pics:RV_HD159062B}
\end{figure*}

\begin{figure*}
\centering
\includegraphics[height=0.3\textwidth]{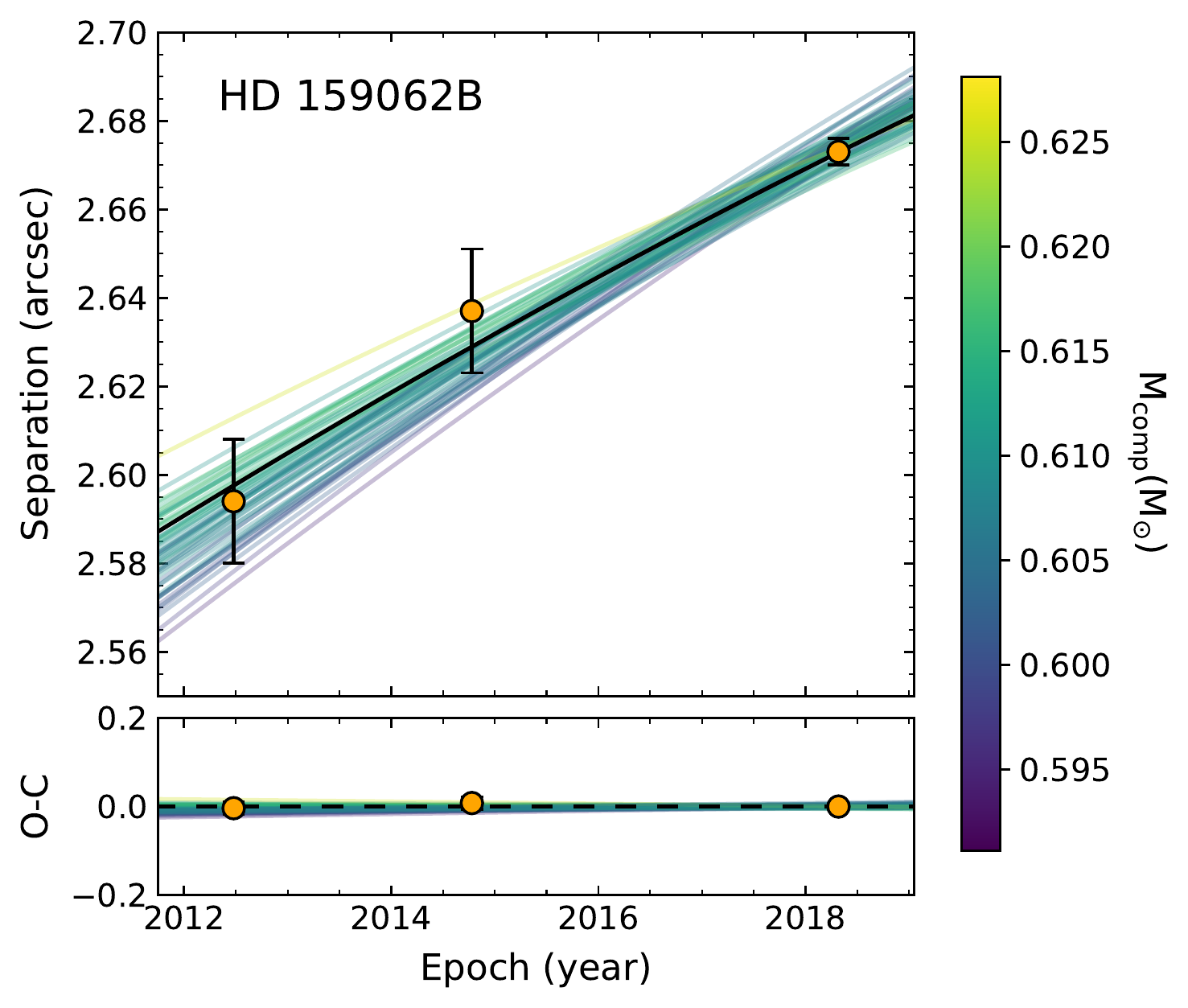}\qquad\qquad
\includegraphics[height=0.3\textwidth]{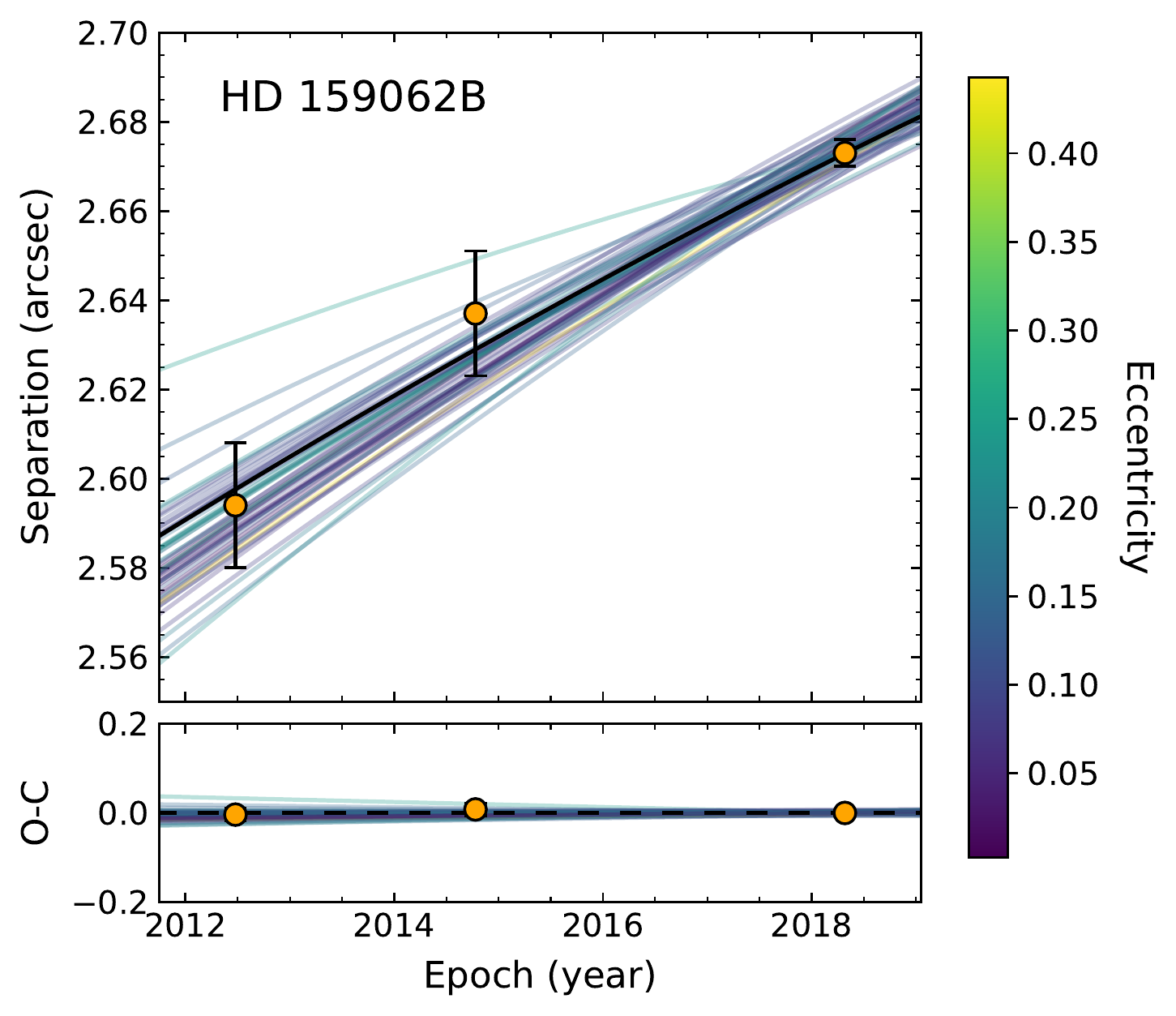}
\includegraphics[height=0.3\textwidth]{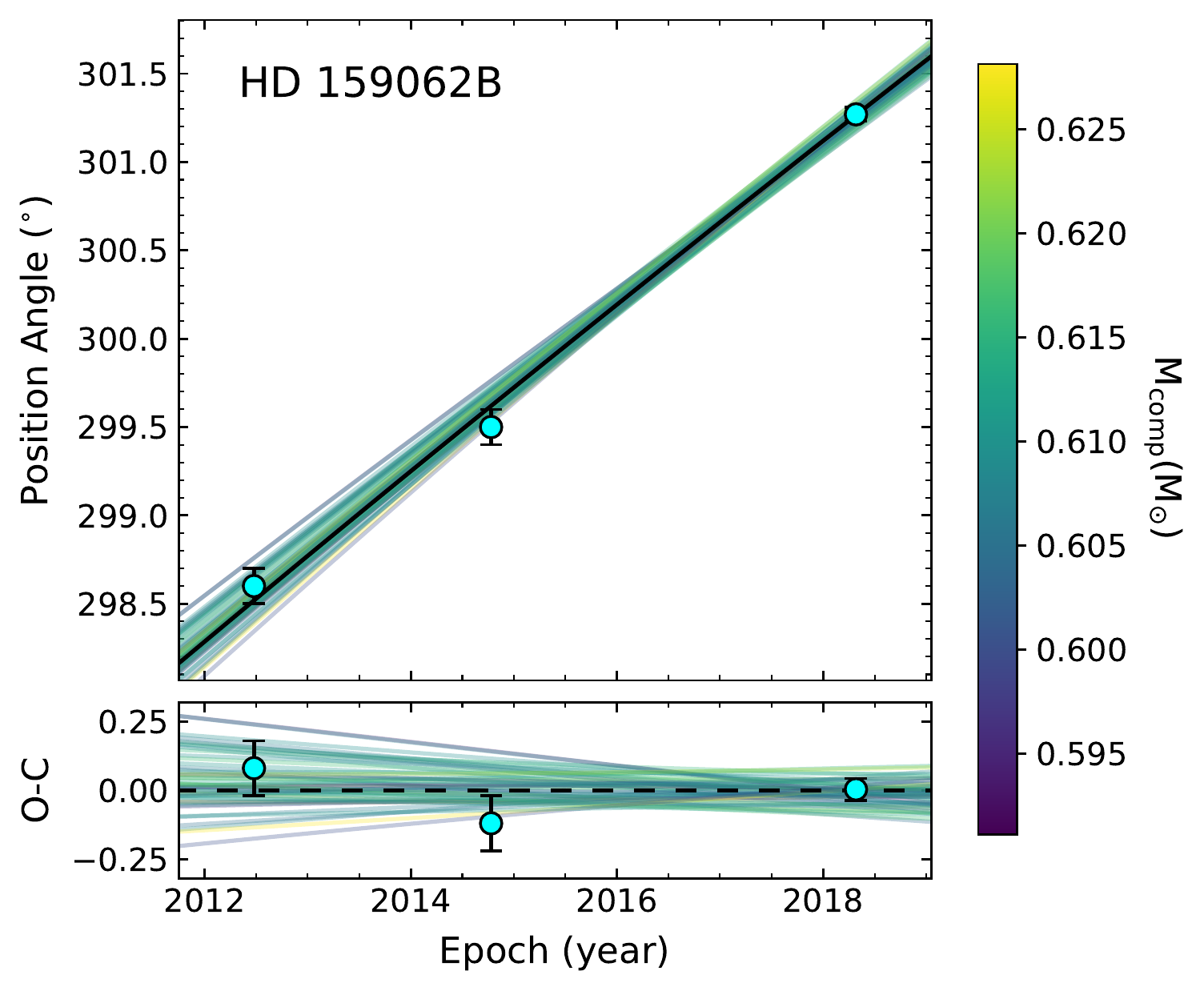} \qquad\qquad
\includegraphics[height=0.3\textwidth]{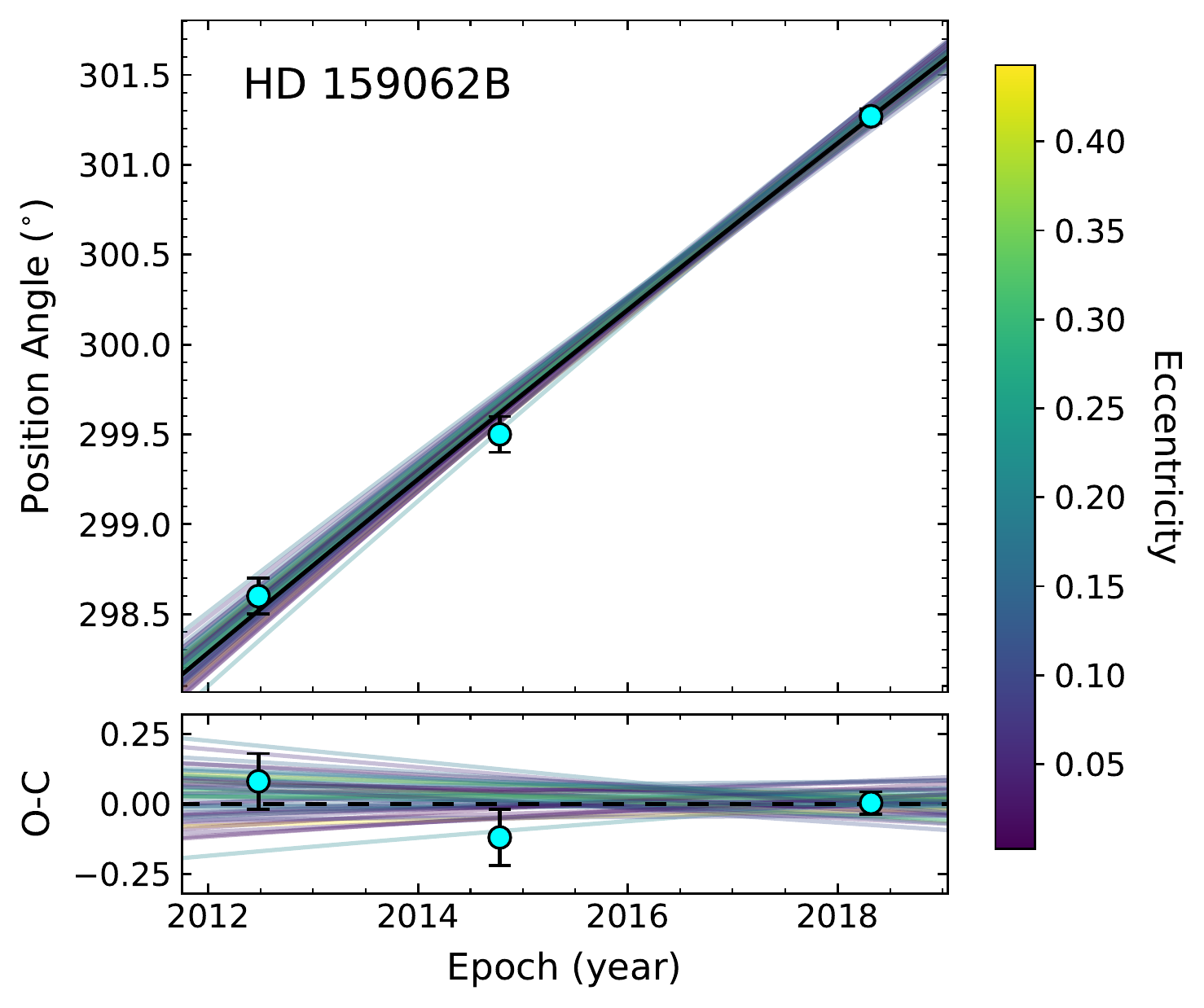}
\includegraphics[height=0.3\textwidth]{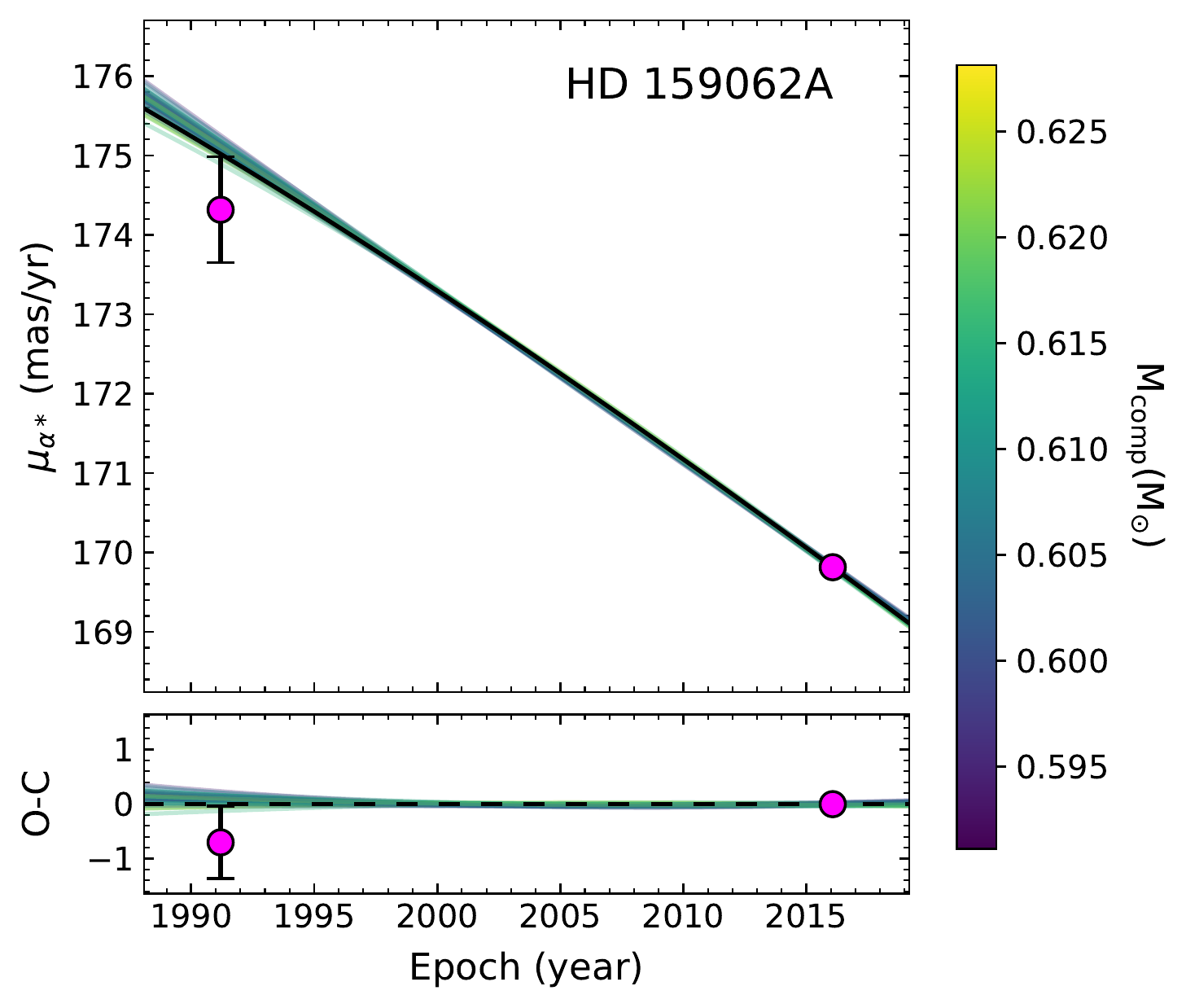}\qquad\qquad
\includegraphics[height=0.3\textwidth]{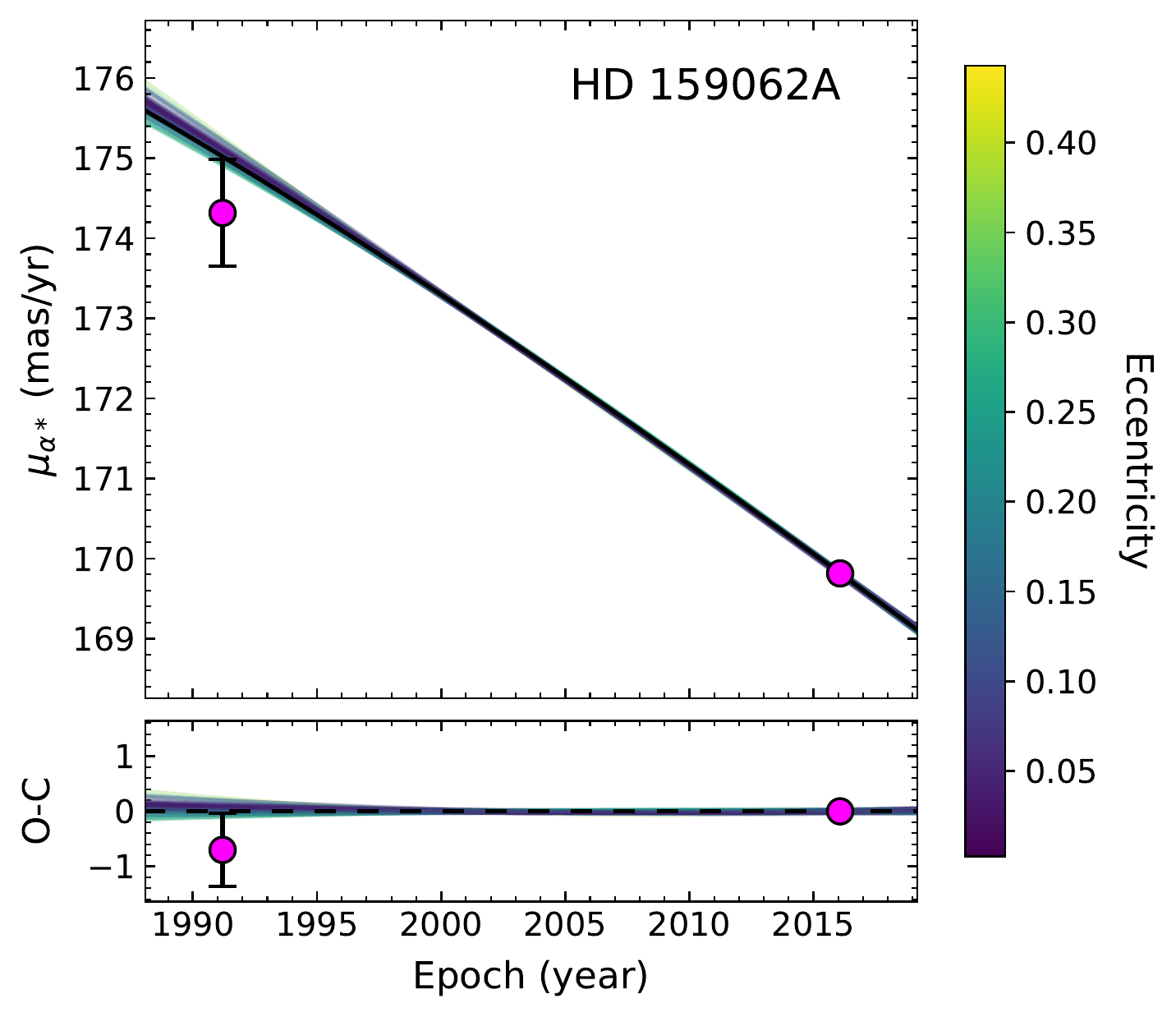}
\includegraphics[height=0.3\textwidth]{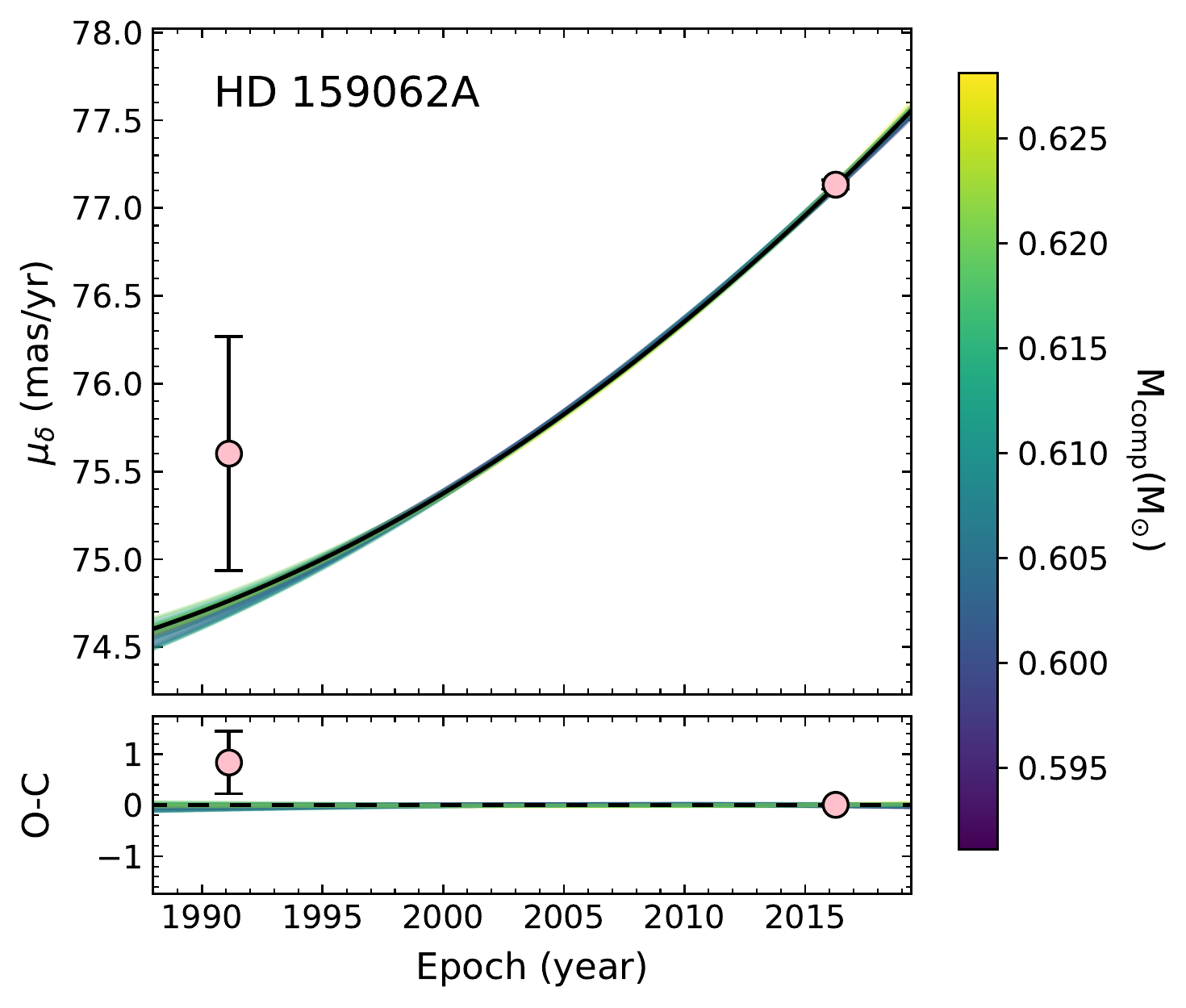} \qquad\qquad
\includegraphics[height=0.3\textwidth]{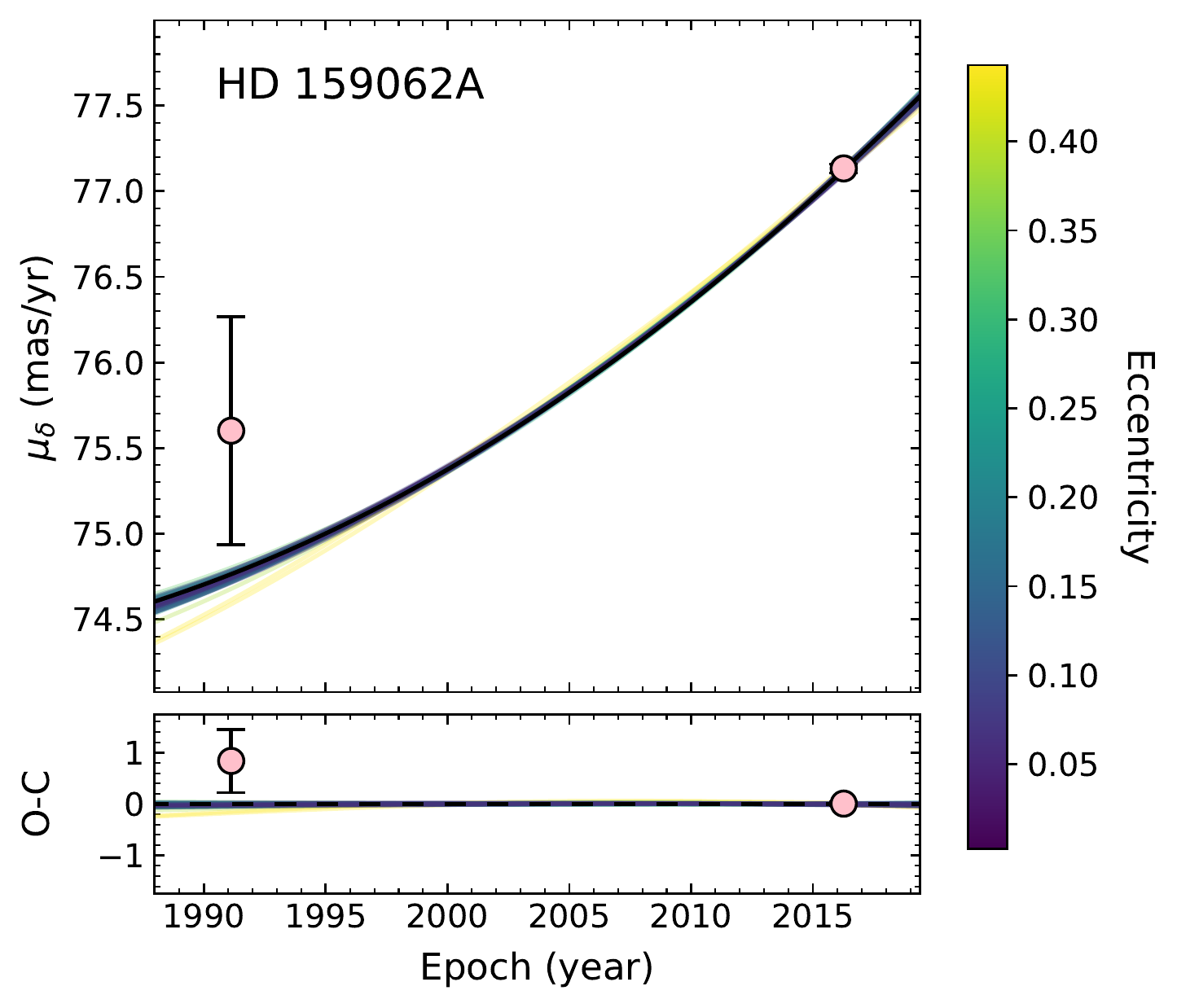}
\caption{Observed and fitted relative and absolute astrometry for the HD~159062AB system.  From top to bottom, the panels show the relative separation and position angle of HD~159062B, and the proper motion of HD~159062A in right ascension and declination.  The thicker black lines represent the best-fit orbit in the MCMC chain while the other 50 lines represent random draws from the chain, color-coded by either companion mass (left panels) or eccentricity (right panels).  The lower insets should the difference between observed and calculated astrometry.  The mean proper motion is used to compute the MCMC chain, but is not shown in the proper motion plots.  This constraint is an integral over the proper motion between the \hipparcos and \gaia epochs.}
\label{pics:relSep_PA_PM_HD159062B}
\end{figure*}

\begin{figure*}
\centering
\includegraphics[width= \textwidth]{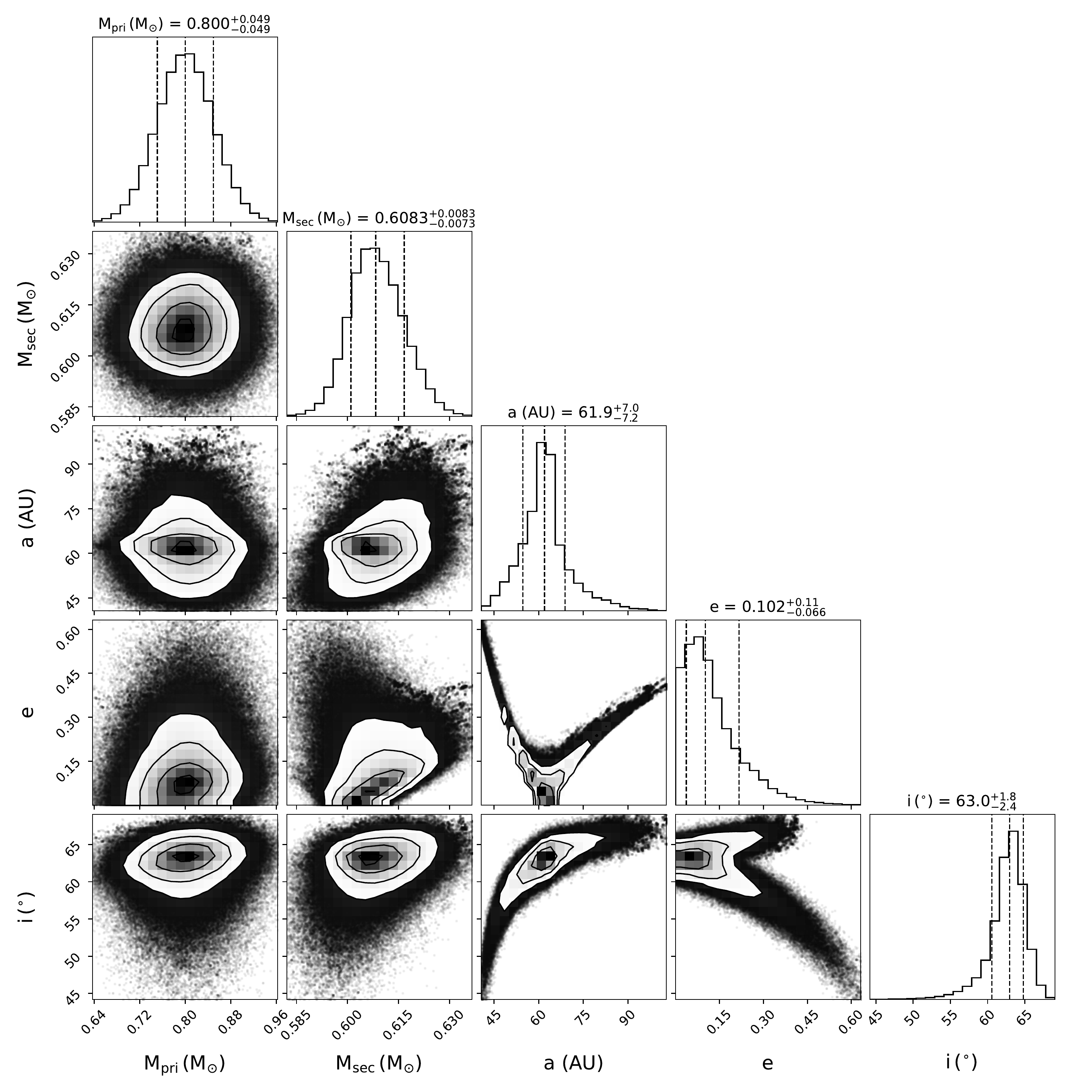}
\caption{Corner plot of the derived parameters including mass of the primary star, mass of the secondary companion, semi-major axis, eccentricity, and inclination. The semimajor axis and eccentricity are highly covariant with one another and, to a lesser degree, with inclination.}
\label{pics:cornerplot_HD159062B}
\end{figure*}

Figure \ref{pics:astrometry_HD159062B} shows the astrometric orbit of HD 159062B. The RVs, both over a full orbit and restricted to the time frame of observations, are displayed in Figure \ref{pics:RV_HD159062B}. The relative separation and position angles of the companion HD 159062B with respect to the host star HD 159062A, and the absolute astrometry of the host star from HGCA are illustrated in Figure \ref{pics:relSep_PA_PM_HD159062B}. Finally, to evaluate the full behavior of the joint posterior distributions, we show a corner plot of the derived parameters of HD 159062B in Figure \ref{pics:cornerplot_HD159062B}.

Our joint orbit fit yields a companion mass of {${0.608}_{-0.0073}^{+0.0083} \Msun$, an orbiting period of $P = {411}_{-70}^{+71}$ years and an eccentricity of ${0.102}_{-0.066}^{+0.11}$} for the white dwarf HD 159062B. 
With our use of HGCA astrometry, we improve H19's constraint on HD 159062B's mass by an order of magnitude.
Our tight constraints on the companion mass and eccentricity help to refine our picture of what the stellar system looks like. This can be used to infer the mass of the progenitor via the initial-final mass relation \citep{Kalirai+Hansen+Kelson+etal_2008}, which could serve as an input on the Barium enrichment of the primary star. We firmly place the system on the long period and low eccentricity tail of the distributions shown in Figure 6 of H19.
\textcolor{black}{This favors barium enrichment via wind accretion and rules out mass transfer by Roche lobe overflow. If the abundance measurement of $[{\rm Ba/Fe}] = 0.4$ dex is accurate, HD 159062 would add to the few known cases of barium enrichment from wind accretion. This new information can be used to guide studies on the local stellar populations and Ba stars. }

\section{Conclusions}
\label{sec:conclusion}

In this paper, we have presented the orbit fitting package \codename, designed to fit one or more Keplerian orbits to any combination of radial velocity, relative astrometry, and absolute astrometry.  \codename achieves high performance by using a combination of a fast eccentric anomaly solver (faster than a standalone call to sine and cosine for a large number of data points), analytic marginalization of parallax and barycenter proper motion, and low-level memory management.  \codename is free and open-source.  It depends on other free packages including {\tt emcee} \citep{Foreman-Mackey+Hogg+Lang+etal_2013}, {\tt ptemcee} \citep{Vousden+Farr+Mandel_2016}, {\tt htof} \citep{Brandt+Michalik+Brandt+etal_2021}, {\tt astropy} \citep{astropy:2013, astropy:2018}, along with {\tt numpy} \citep{numpy1, numpy2} and {\tt scipy} \citep{2020SciPy-NMeth}. \codename may be downloaded from {\tt github} and installed by {\tt pip}.  We have tested its installation and operation on {Windows,} Mac and Linux machines.

We have demonstrated \codename on the system HD~159062, a white dwarf-main sequence binary discovered by \cite{Hirsch_2019}.  Those authors found a white dwarf mass of $0.65_{-0.04}^{+0.12}~M_\odot$, and a semimajor axis and eccentricity that were compatible with a close pericenter passage.  By combining absolute astrometry with the radial velocity and relative astrometry used by \cite{Hirsch_2019}, we improve the precision of the mass by an order of magnitude, to $0.617_{-0.012}^{+0.013}~M_\odot$.  We exclude orbits with a close pericenter passage, establishing that the binary was a case of wind accretion system in its past.

\codename supports a variety of applications beyond what we have demonstrated here with HD~159062.  It can account for multiple companions, it appropriately treats epoch astrometry from \hipparcos and/or \gaia, and it can account for a companion proper motion measured by \gaia.  \codename is also designed to produce publication-ready plots.  \codename may be configured with a single {\tt .ini} file.  We plan to keep \codename current, particularly when new \gaia data releases enable substantial improvements in astrometric precision.  

\acknowledgments{{We thank an anonymous referee for many helpful comments, suggestions that improved the paper and the structure and usability of the code. G.~M.~B. is supported by the National Science Foundation (NSF) Graduate Research Fellowship under grant no. 1650114.}}

\clearpage
\appendix
\label{sec:appendix}

Here we derive the multivariate Gaussian that describes the parallax and barycenter proper motion at fixed values of the other orbital parameters.  Marginalizing over this multivariate Gaussian is equivalent to replacing the parallax and proper motion with their mean (i.e.~best-fit) values and multiplying by the square root of the determinant of the covariance matrix.

We first write out the components of $\chi^2$ that involve the parallax $\varpi$ and the proper motion of the system barycenter $\bm{\mu}$.  We assume here that we have a measurement of the secondary star's proper motion from \gaia; this measurement has a proper motion $\bm{\mu}_{G, \rm o, B}$ and an inverse covariance matrix ${\bf C}_{G, \rm B}^{-1}$.  Measurements of the primary star use a suffix A, and proper motion model values refer to the proper motion of the primary star unless they carry the subscript B.  If the secondary lacks a proper motion measurement in \gaia (as is the case for the companions in \citealt{Brandt+Dupuy+Bowler_2019} and \citealt{Dupuy+Brandt+Kratter+etal_2019}), then we set ${\bf C}_{G, \rm B}^{-1} = 0$.  We also take model proper motions to be in physical units (e.g.~AU\,yr$^{-1}$), so that multiplying by parallax gives proper motions in angular units.

\begin{align}
\chi^2 = & \frac{\left(\varpi - \varpi_{\it Gaia}\right)^2}{\sigma^2_{\varpi,\it Gaia}} + \sum_{k=1}^{N_{\rm ast}} \frac{\left( \rho_k-\varpi\rho\left[t_k\right] \right)^2}{(1 - c^2_{\rho\theta,k}) \sigma^2 [\rho_k]} - 2 \sum_{k=1}^{N_{\rm ast}} \frac{c_{\rho\theta,k}\lfloor\theta_k - \theta[t_k]\rfloor \left( \rho_k-\varpi\rho\left[t_k\right] \right)}{(1 - c^2_{\rho\theta,k})\sigma [\rho_k]\sigma [\theta_k]} \nonumber \\
    &
    +\left( {\bm \mu_{H, \rm o, A}} - \overline{\bm \mu} - \varpi \bm{\mu}_{H} \right)^T {\bf C}_{H, \rm A}^{-1} \left( {\bm \mu_{H, \rm o, A}} - \overline{\bm \mu} - \varpi \bm{\mu}_{H} \right) \nonumber \\
    &+\left( {\bm \mu_{HG, \rm o, A}} - \overline{\bm \mu} - \varpi \bm{\mu}_{HG} \right)^T {\bf C}_{HG, \rm A}^{-1} \left( {\bm \mu_{HG, \rm o, A}} - \overline{\bm \mu} - \varpi \bm{\mu}_{HG} \right) \nonumber \\
    &
    +\left( {\bm \mu_{G,\rm o, A}} - \overline{\bm \mu} - \varpi \bm{\mu}_{G} \right)^T {\bf C}_{G, \rm A}^{-1} \left( {\bm \mu_{G, \rm o, A}} - \overline{\bm \mu} - \varpi \bm{\mu}_{G} \right) \nonumber \\
    &
    + \left( {\bm \mu_{G, \rm o, B}} - \overline{\bm \mu} - \varpi \bm{\mu}_{G,\rm B} \right)^T {\bf C}_{G, \rm B}^{-1} \left( {\bm \mu_{G, \rm o, B}} - \overline{\bm \mu} - \varpi \bm{\mu}_{G,\rm B} \right) 
    \label{eq:chisq_marginalize}
\end{align}
We can minimize $\chi^2$ with respect to parallax and proper motion of the barycenter by solving
\begin{align}
    {\bf M} 
    \begin{bmatrix}
    \varpi \\
    \overline{\mu}_{\alpha*} \\
    \overline{\mu}_{\delta}
    \end{bmatrix}
    =
    {\bf b}
    \label{eq:parsolve}
\end{align}
where $M$ is the symmetric matrix with elements
\begin{align}
    M_{\varpi \varpi} &= \bm{\mu}_{H}^T {\bf C}_{H, \rm A}^{-1} \bm{\mu}_{H} 
    + \bm{\mu}_{HG}^T {\bf C}_{HG, \rm A}^{-1} \bm{\mu}_{HG} 
    + \bm{\mu}_{G}^T {\bf C}_{G, \rm A}^{-1} \bm{\mu}_{G}
    + \bm{\mu}_{G,\rm B}^T {\bf C}_{G, \rm B}^{-1} \bm{\mu}_{G,\rm B} 
    + \frac{1}{\sigma^2_{\varpi,\it Gaia}} + \sum_{k=1}^{N_{\rm ast}} \frac{\rho^2\left[t_k\right]}{(1 - c^2_{\rho\theta,k})\sigma^2 [\rho_k]} \\
    M_{\varpi \alpha} &= \bm{\mu}_{H} \cdot \begin{bmatrix} C^{-1}_{H,{\rm A},\alpha\alpha} \\ C^{-1}_{H,{\rm A},\alpha\delta} \end{bmatrix} 
    + \bm{\mu}_{HG} \cdot \begin{bmatrix} C^{-1}_{HG,{\rm A},\alpha\alpha} \\ C^{-1}_{HG,{\rm A},\alpha\delta} \end{bmatrix}  
    + \bm{\mu}_{G} \cdot \begin{bmatrix} C^{-1}_{G,{\rm A},\alpha\alpha} \\ C^{-1}_{G,{\rm A},\alpha\delta} \end{bmatrix} 
    + \bm{\mu}_{G,\rm B} \cdot \begin{bmatrix} C^{-1}_{G,{\rm B},\alpha\alpha} \\ C^{-1}_{G,{\rm B},\alpha\delta} \end{bmatrix} \\
    M_{\varpi \delta} &= \bm{\mu}_{H} \cdot \begin{bmatrix} C^{-1}_{H,{\rm A},\alpha\delta} \\ C^{-1}_{H,{\rm A},\delta\delta} \end{bmatrix} 
    + \bm{\mu}_{HG} \cdot \begin{bmatrix} C^{-1}_{HG,{\rm A},\alpha\delta} \\ C^{-1}_{HG,{\rm A},\delta\delta} \end{bmatrix}  
    + \bm{\mu}_{G} \cdot \begin{bmatrix} C^{-1}_{G,{\rm A},\alpha\delta} \\ C^{-1}_{G,{\rm A},\delta\delta} \end{bmatrix} 
    + \bm{\mu}_{G,\rm B} \cdot \begin{bmatrix} C^{-1}_{G,{\rm B},\alpha\delta} \\ C^{-1}_{G,{\rm B},\delta\delta} \end{bmatrix} \\
    M_{\alpha\alpha} &= C^{-1}_{H,{\rm A},\alpha\alpha} + C^{-1}_{HG,{\rm A},\alpha\alpha} + C^{-1}_{G,{\rm A},\alpha\alpha} + C^{-1}_{G,{\rm B},\alpha\alpha} \\
    M_{\delta\delta} &= C^{-1}_{H,{\rm A},\delta\delta} + C^{-1}_{HG,{\rm A},\delta\delta} + C^{-1}_{G,{\rm A},\delta\delta} + C^{-1}_{G,{\rm B},\delta\delta} \\
    M_{\alpha\delta} &= C^{-1}_{H,{\rm A},\alpha\delta} + C^{-1}_{HG,{\rm A},\alpha\delta} + C^{-1}_{G,{\rm A},\alpha\delta} + C^{-1}_{G,{\rm B},\alpha\delta}
\end{align}
and ${\bf b}$ is given by
\begin{align}
    b_{\varpi} &= \bm{\mu}_{H,\rm o}^T {\bf C}_{H, \rm A}^{-1} \bm{\mu}_{H} 
    + \bm{\mu}_{HG,\rm o}^T {\bf C}_{HG, \rm A}^{-1} \bm{\mu}_{HG} 
    + \bm{\mu}_{G,\rm o}^T {\bf C}_{G, \rm A}^{-1} \bm{\mu}_{G}
    + \bm{\mu}_{G,\rm o, B}^T {\bf C}_{G, \rm B}^{-1} \bm{\mu}_{G,\rm B} + \frac{\varpi_{\it Gaia}}{\sigma^2_{\varpi,\it Gaia}}
    \nonumber \\ &\qquad\qquad + \sum_{k=1}^{N_{\rm ast}} \frac{\rho_k \rho[t_k]}{(1 - c^2_{\rho \theta,k})\sigma^2[\rho_k]} - \sum_{k=1}^{N_{\rm ast}} \frac{c_{\rho \theta,k}\rho[t_k]\lfloor\theta_k - \theta[t_k]\rfloor}{(1 - c^2_{\rho \theta,k})\sigma[\rho_k]\sigma[\theta_k]}\\
    b_{\alpha} &= \bm{\mu}_{H,\rm o} \cdot \begin{bmatrix} C^{-1}_{H,{\rm A},\alpha\alpha} \\ C^{-1}_{H,{\rm A},\alpha\delta} \end{bmatrix} 
    + \bm{\mu}_{HG,\rm o} \cdot \begin{bmatrix} C^{-1}_{HG,{\rm A},\alpha\alpha} \\ C^{-1}_{HG,{\rm A},\alpha\delta} \end{bmatrix}  
    + \bm{\mu}_{G,\rm o} \cdot \begin{bmatrix} C^{-1}_{G,{\rm A},\alpha\alpha} \\ C^{-1}_{G,{\rm A},\alpha\delta} \end{bmatrix} 
    + 
    \bm{\mu}_{G,\rm o, B} \cdot \begin{bmatrix} C^{-1}_{G,{\rm B},\alpha\alpha} \\ C^{-1}_{G,{\rm B},\alpha\delta} \end{bmatrix} \\
    b_{\delta} &= \bm{\mu}_{H,\rm o} \cdot \begin{bmatrix} C^{-1}_{H,{\rm A},\alpha\delta} \\ C^{-1}_{H,{\rm A},\delta\delta} \end{bmatrix} 
    + \bm{\mu}_{HG,\rm o} \cdot \begin{bmatrix} C^{-1}_{HG,{\rm A},\alpha\delta} \\ C^{-1}_{HG,{\rm A},\delta\delta} \end{bmatrix}  
    + \bm{\mu}_{G,\rm o} \cdot \begin{bmatrix} C^{-1}_{G,{\rm A},\alpha\delta} \\ C^{-1}_{G,{\rm A},\delta\delta} \end{bmatrix} 
    + 
    \bm{\mu}_{G,\rm o, B} \cdot \begin{bmatrix} C^{-1}_{G,{\rm B},\alpha\delta} \\ C^{-1}_{G,{\rm B},\delta\delta} \end{bmatrix}. 
\end{align}
The covariance matrix of $\varpi$ and the two components of $\overline{\bm{\mu}}$ is the inverse of ${\bf M}$.  Integrating, or marginalizing, over $\overline{\bm{\mu}}$ and $\varpi$ is equivalent to replacing $\overline{\bm{\mu}}$ and $\varpi$ in Equation \eqref{eq:chisq_marginalize} with the solutions of Equation \eqref{eq:parsolve} and dividing by the square root of the determinant of ${\bf M}$ (multiplying by the square root of the determinant of the covariance matrix ${\bf M}^{-1}$).

\bibliographystyle{apj_eprint}
\bibliography{refs.bib}

\end{document}